\documentclass[prd, twocolumn, nofootinbib, floatfix]{revtex4}

\usepackage{amsmath}
\usepackage{amsbsy}

\input epsf
\usepackage{amsmath,amssymb,subfigure}
\usepackage{graphicx}
\usepackage{epsfig}
\usepackage{color}

\newcommand{\be}{\begin{equation}}
\newcommand{\ee}{\end{equation}}
\newcommand{\gscr}{{\mathcal{G}}}
\newcommand{\vscr}{{\mathcal{V}}}

\newcommand{\beq}{\begin{equation}}
\newcommand{\eeq}{\end{equation}}
\newcommand{\beqa}{\begin{eqnarray}}
\newcommand{\eeqa}{\end{eqnarray}}
\newcommand{\om}{\Omega_m}

\newcommand{\ls}{\mathrel{\raise0.27ex\hbox{$<$}\kern-0.70em \lower0.71ex\hbox{{
$\scriptstyle \sim$}}}}

\begin{document} 

\title{Confronting General Relativity with Further Cosmological Data} 
\author{Scott F.\ Daniel$^1$ and Eric V.\ Linder$^{1,2,3}$
} 
\affiliation{$^1$Institute for the Early Universe, Ewha Womans University, 
Seoul, Korea\\ 
$^2$Lawrence Berkeley National Laboratory, Berkeley, CA, USA\\ 
$^3$Berkeley Center for Cosmological Physics, University of California, 
Berkeley, CA, USA 
}

\date{\today}

\begin{abstract} 
Deviations from general relativity in order to explain cosmic acceleration 
generically have both time and scale dependent signatures in cosmological 
data.  We extend our previous work by investigating model independent 
gravitational deviations in bins of redshift and length scale, by 
incorporating further cosmological probes such as temperature-galaxy and 
galaxy-galaxy cross-correlations, and by examining correlations between 
deviations.  Markov Chain Monte Carlo likelihood analysis of the model 
independent parameters fitting current data indicates that at low redshift 
general relativity deviates from the best fit at the 99\% confidence level. 
We trace this to two different properties of the CFHTLS weak lensing 
data set and demonstrate that COSMOS weak lensing data does not show 
such deviation.  Upcoming galaxy survey data will greatly improve the 
ability to test time and scale dependent extensions to gravity and we 
calculate the constraints that the BigBOSS galaxy redshift survey could 
enable. 
\end{abstract} 

\maketitle

\section{Introduction \label{sec:intro}}

Gravitation is the key force governing the expansion and evolution 
of the universe.  The unexpected observations of cosmic acceleration 
may indicate that some aspects of this fundamental force remain a 
mystery.  General relativity is a hugely successful theory of gravity 
over the ranges it has been tested, but we should continue to test it in 
greater detail in regions, such as on cosmic scales, where it has not 
been sufficiently probed. 

Since it is not clear what form deviations from general relativity 
(GR) may take, it is useful not only to adopt specific models extending 
GR but also to consider model independent approaches.  These generally 
parameterize the relation between the metric potentials, the relation 
between the potentials and the matter density, or similar forms.  A 
translation table between many of the most common conventions was 
provided in \cite{gr1}.  

The effects giving rise to cosmic acceleration must take place on 
the largest length scales, but general relativity is known to be 
highly accurate on small scales (solar system and laboratory), so 
the deviations must have scale dependence.  This can either be innate 
(from the scale dependence in the Poisson equation), or explicit. 
Similarly, conditions in the early universe such as during primordial 
nucleosynthesis or recombination can be well explained within GR, 
and acceleration is a recent phenomenon, so the deviation from GR 
should also be time dependent. 

In this article we broaden consideration of the deviation parametrization, 
and convert to more observationally direct variables than used in 
\cite{gr1}.  In Sec.~\ref{sec:par} 
we examine some possible time and space dependencies and examine the 
correlation between the deviation variables.  By adopting a model 
independent, binned formalism we avoid putting in ad hoc assumptions 
about the form of the deviation, letting the data determine the results.  
We consider different data types probing the matter density distribution 
in Sec.~\ref{sec:tggg}, going beyond the cosmic microwave background (CMB) 
perturbations, Type Ia supernova distance-redshift relation, and weak 
gravitational lensing used in \cite{gr1}.  Prospects for further 
improvements in constraints from future data are investigated in 
Sec.~\ref{sec:fut}.

\section{Constraining Deviations of Gravity \label{sec:par}} 

The relations between the two metric potentials (often called the 
gravitational slip), the matter density and velocity fields 
(continuity equation), the matter density and a metric potential 
(Poisson equation), and velocity field and the other metric 
potential (Euler equation) form a system of equations describing 
the spacetime and its contents.  Modifications to gravity adjust 
these interrelations and so one can parameterize these theories by 
inserting time and space dependent functions in the usual GR 
relations. 

One example is to define the gravitational slip as 
\beq 
\psi=[1+\varpi(a,k)]\,\phi\,, \label{eq:varpidef}
\eeq 
where the metric is given in conformal Newtonian gauge through 
\beq 
ds^2=a^2[-(1+2\psi)\,d\tau^2+(1-2\phi)\,d\vec x^2]\,, 
\eeq 
and $a$ is the scale factor, $k$ the wavenumber, $\tau$ the conformal 
time, and $x$ 
the spatial coordinate.  Preserving stress energy conservation, 
and so the continuity and Euler equations, we are left with needing the 
Poisson equation, modified to 
\beq 
-k^2\phi=4\pi G_N  a^2\bar\rho_m\Delta_m\times \mu(a,k)\,, \label{eq:mudef} 
\eeq 
where $G_N$ is Newton's constant, $\bar\rho_m$ is the homogeneous part of 
the matter density 
and $\Delta_m$ the perturbed part written in gauge-invariant form, i.e.
\beq 
\Delta_m\equiv\delta_m+\frac{3\mathcal{H}\theta_m}{k^2} 
\eeq 
in the notation of \cite{Ma:1995ey}, where $\theta_m$ is the 
velocity perturbation and $\mathcal{H}$ is the conformal Hubble parameter. 

These two functions, $\varpi$ and $\mu$, were used in \cite{gr1} 
and several other papers, and are equivalent to many other 
parametrizations as detailed in the translation table of \cite{gr1}. 
In this paper, we will present a few further results using these 
variables, but the bulk of the paper will use ``decorrelated'' 
parameters based on these.

\subsection{$\varpi$ and $\mu$ as Variables \label{sec:vmu}} 

One of the main focuses in \cite{gr1} was to test consistency with 
GR.  For this, only one of $\varpi$ or $\mu$ were varied at a time. 
Since a shift in $\varpi$ could be compensated by a corresponding 
shift in $\mu$ (see Fig.~2 of \cite{gr1}, or the degeneracy line in 
their Fig.~7) to preserve the observational agreement, the intent 
of varying one at a time was to make it more difficult to achieve 
agreement with GR and hence provide a more conservative result. 
Despite this ``handicapping'', agreement with GR indeed occurred. 

\begin{figure}[!t]
\includegraphics[angle=-90,width=\columnwidth]{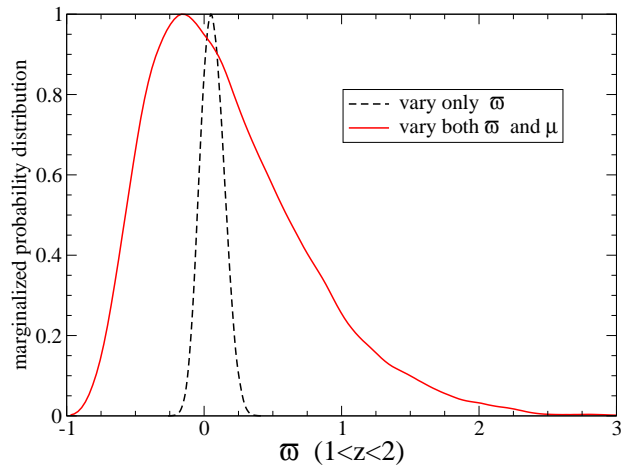}
\caption{
1D marginalized probability of the post-GR parameter $\varpi$ in 
the redshift bin $1<z<2$.  The narrower, dashed (black) distribution fixes 
the other post-GR parameter $\mu=1$, making it more difficult 
to fit the data (consistent with GR) by compensating one parameter 
with another.  The wider, 
solid (red) distribution includes a simultaneous fit for $\mu$.  All other 
cosmological parameters, including $\varpi$ and $\mu$ in the lower 
and higher redshift bins, are marginalized over.  
}
\label{fig:1Dvwithmu}
\end{figure}

In Fig.~\ref{fig:1Dvwithmu} we show what happens when $\mu$ is 
allowed to vary simultaneously with $\varpi$, treating both functions 
as being composed of constant values within each of three redshift bins 
($z<1$, $1<z<2$, and $2<z<9$, with $z>9$ fixed to GR).  
To compare to \cite{gr1}, we use data constraints from 
WMAP 5 year CMB \cite{Dunkley:2008ie, Nolta:2008ih, Hinshaw:2008kr}, 
Union2 supernova distances \cite{amanullah}, and COSMOS weak lensing 
\cite{Massey:2007gh} data sets.  
The narrow 1D distribution of $\varpi$ recreates the fixed $\mu$ case 
of the middle panel of Fig.~5 of \cite{gr1}, while the wider distribution 
shows the results for $\varpi$ when also fitting $\mu$.  The 68\% cl 
range increases by approximately a factor of 6. 
The other redshift bins behave similarly.  Thus, the accuracy of 
measurement of the deviations $\varpi$ (and $\mu$) is not particularly 
tight.

\begin{figure}[!t]
\includegraphics[angle=-90,width=\columnwidth]{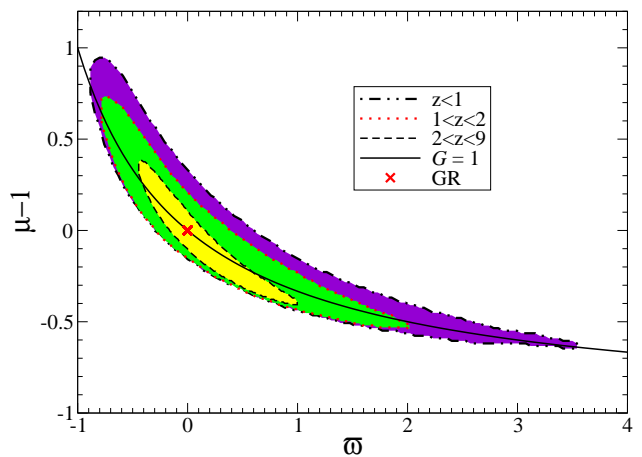} 
\caption{
2D joint probability contours at 95\% cl between the post-GR functions 
$\varpi$ and $\mu$, for three independent redshift bins.
Values within the 
redshift bins are consistent with each other and with GR 
(denoted by the cross at (0,0)). 
The solid, black curve, motivated by a modified Poisson equation, 
closely follows the degeneracy direction, and suggests a more 
insightful parametrization using variables along and perpendicular 
to the curve. 
}
\label{fig:muvarpi6}
\end{figure}

The degeneracy between the two post-GR functions is clearly seen in 
the 2D probability distributions of Fig.~\ref{fig:muvarpi6}.  The 
banana shape discussed in \cite{gr1} persists here, even though we 
use independent bins of redshift rather than the $a^3$ functional 
dependence assumed in their Fig.~7.  The solid black curve shows a 
theoretically motivated compensation relation largely responsible 
for the degeneracy.

We will be able to improve the constraints, and our understanding of 
the observational leverage on modified gravity, by ``trading'' precision 
on one combination of $\varpi$ and $\mu$ for that of another combination.  
This basically corresponds to choosing variables along and perpendicular 
to the main degeneracy direction, as we now discuss.

\subsection{Separating Parameter Effects: $\gscr$ and $\vscr$ \label{sec:geff}} 

Several types of observables are predominantly sensitive to the 
sum of the potentials, e.g.\ the integrated Sachs-Wolfe (ISW) effect and 
gravitational lensing.  Writing the Poisson equation in terms of 
such a sum yields 
\beq 
-k^2(\phi+\psi)=8\pi G_N a^2\bar\rho_m\Delta_m \times \gscr\,, \label{eq:phipsi} 
\eeq 
where Eqs.~(\ref{eq:varpidef}) and (\ref{eq:mudef}) show that 
$\gscr=\mu\,(2+\varpi)/2$.  It is not surprising therefore that 
the confidence contours in the $\mu$-$\varpi$ plane are banana 
shaped with strong curvature. 

The main degeneracy curve illustrated in Fig.~\ref{fig:muvarpi6} 
is precisely the combination entering $\gscr$.  Therefore it makes 
sense to switch variables to use this combination as one parameter.  We can 
also use Eqs.~(\ref{eq:phipsi}) and (\ref{eq:mudef}) to define a 
Poisson-like equation for $\psi$; here we write all three equations 
together to show the parallelism: 
\beqa 
-k^2(\phi+\psi)&=&8\pi G_N a^2\bar\rho_m\Delta_m \times \gscr \\ 
-k^2\phi&=&4\pi G_N  a^2\bar\rho_m\Delta_m \times \mu \\ 
-k^2\psi&=&4\pi G_N  a^2\bar\rho_m\Delta_m \times \vscr\,. \label{eq:vscrdef} 
\eeqa 
The parameter $\vscr$ is precisely the parameter identified as mostly 
sensitive to growth of structure in \cite{gr1} (there called $\Sigma$; 
note that \cite{song} earlier noted this set to be of interest, calling 
$\vscr$ as $\mu$, and $\gscr$ as $\Sigma$; \cite{Zhao:2010dz} also explored 
this later).  It is also closely related 
to the growth index parameter $\gamma$ \cite{groexp, Linder:2007hg}. 

The new post-GR functions are related to the old ones via 
\beqa 
\gscr&=&\mu\,\frac{2+\varpi}{2} \qquad ; \qquad \mu=2\gscr-\vscr \\ 
\vscr&=&\mu\,(1+\varpi) \quad ; \qquad \varpi=\frac{2\vscr-2\gscr}{2\gscr-\vscr} \,. \label{eq:trans} 
\eeqa 
We will see that the new functions are substantially decorrelated 
from each other, producing more independent constraints when using the 
observational data.  
(The symbol $\gscr$ is meant to evoke an effective Newton's constant 
in the total Poisson equation; $\vscr$ recalls the ``velocity'' equation 
arising from the relation between the potential $\psi$ and the matter 
velocity field, central to growth of structure.) 

One expects that the integrated Sachs-Wolfe effect 
and, in large part, weak gravitational 
lensing data will mostly constrain $\gscr$ (i.e.\ across the degeneracy 
direction seen in Fig.~\ref{fig:muvarpi6}) and have little leverage 
on $\vscr$ (i.e.\ along the degeneracy direction seen in 
Fig.~\ref{fig:muvarpi6}).  Probes that involve growth, such as 
weak gravitational lensing and the cross-correlation between the CMB 
and the galaxy density field, to some extent, and the galaxy-galaxy 
density power spectrum, should place some constraint on $\vscr$. 
From Fig.~\ref{fig:muvarpi6} one expects that current weak lensing 
data will not be that strong, however, so we will also investigate 
the role of current density field data in Sec.~\ref{sec:tggg}.  Future 
galaxy survey data should tighten the constraints further; 
see Sec.~\ref{sec:fut} for further discussion.

\subsection{Redshift and Scale Dependence \label{sec:zkdep}} 

The post-GR functions will generally be functions of both time 
(redshift) and length scale (wavenumber).  We do not necessarily want 
to assume a particular functional form, so we begin by allowing the 
values of $\gscr$ and $\vscr$ to take arbitrary values within 
independent bins of redshift $z$ and wavemode $k$ (also see early 
work by \cite{zhao09}). 

If we examine the redshift dependence of $\gscr$, using no 
scale dependence initially, we find that the values of $\gscr$ in 
different bins are positively correlated.  We consider two 
independent redshift bins, with $0<z<1$ and $1<z<2$.  For $z>2$ 
we assume the GR values.  The 
characteristics discussed below do not change if we add a third 
bin at $2<z<9$, but the remainder of this paper uses two bins. 
The degeneracy direction 
between $\gscr(0<z<1)$ and $\gscr(1<z<2)$ corresponds roughly to a 
dependence on scale factor $\gscr(a)\sim a^1$, at least for $z<2$.  
The function $\vscr$ shows a negative correlation between $\vscr(0<z<1)$ 
and $\vscr(1<z<2)$, such that they roughly compensate each other: 
$\vscr(0<z<1)-1\approx -[\vscr(1<z<2)-1]$. 

Regarding degeneracies with other cosmological parameters, there 
is little correlation except with the mass fluctuation amplitude 
$\sigma_8$.  This accords with the 
principal influence of $\vscr$ and $\gscr$ being on growth 
of scalar perturbations, especially at late times.  The main effect 
is a positive correlation between $\vscr(1<z<2)$ and $\sigma_8$; recall 
that $\vscr$ is the post-GR parameter most strongly entering into the 
growth of 
$\Delta_m$.  That the higher $z$ bin of $\vscr$ is most correlated 
follows from growth being cumulative, so the higher redshift bin has 
a longer lever arm of influence to imprint the effects of gravitational 
modifications. 
We also find a slight negative correlation between
$\gscr$ and $\sigma_8$.
This is related to the weak lensing data, which involves the sum of the 
potentials as well as the growth (see the 
discussion at the end of Section \ref{sec:tggg}). 
For higher $\gscr$, lower values of
$\sigma_8$ will produce comparable lensing potentials.  
Thus, larger $\gscr$ does not cause $\sigma_8$ to decrease per se 
(the way larger $\vscr$ amplifies growth),
rather it brings lower values of $\sigma_8$ into better agreement
with the data.

Now considering scale dependence, we introduce two bins in wavenumber $k$, 
running from $k=10^{-4}-10^{-2}~\text{Mpc}^{-1}$ 
and $k>10^{-2}~\text{Mpc}^{-1}$.  The 
low $k$ range represents the large scales from roughly Hubble scale 
to matter-radiation equality horizon scale, and the high $k$ range 
corresponds to scales roughly over which non-CMB probes have leverage. 
For example we expect that the matter power spectrum (including weak 
lensing) would mostly constrain the second bin. 
Thus in total we fit for 8 post-GR parameters: $\gscr$ and $\vscr$ 
values, each in 2 bins of $z$ and 2 bins of $k$.

\begin{figure*}[!t]
\subfigure[]{\includegraphics[angle=-90,width=\columnwidth]{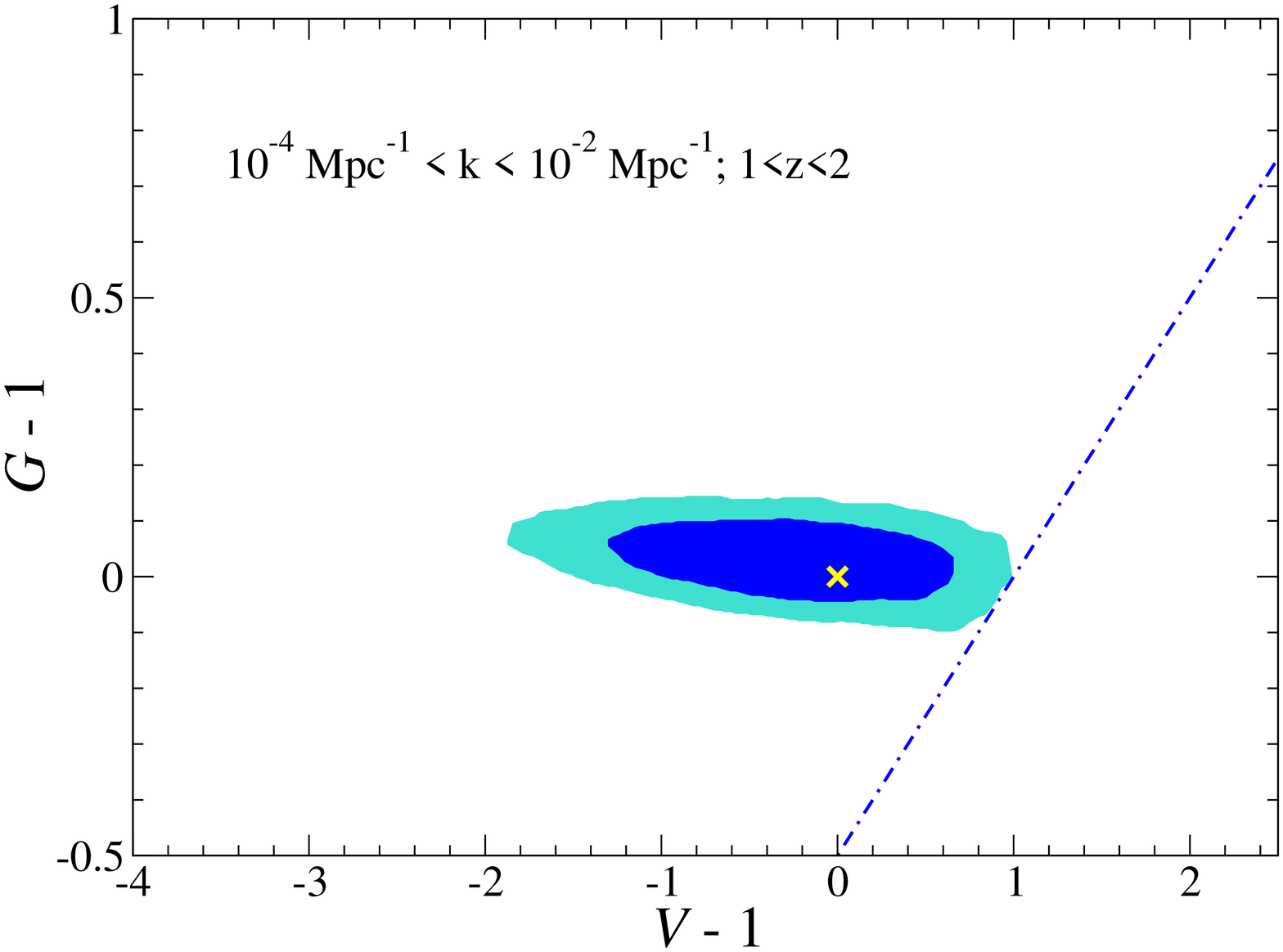}
\label{fig:nogbin1}}
\subfigure[]{\includegraphics[angle=-90,width=\columnwidth]{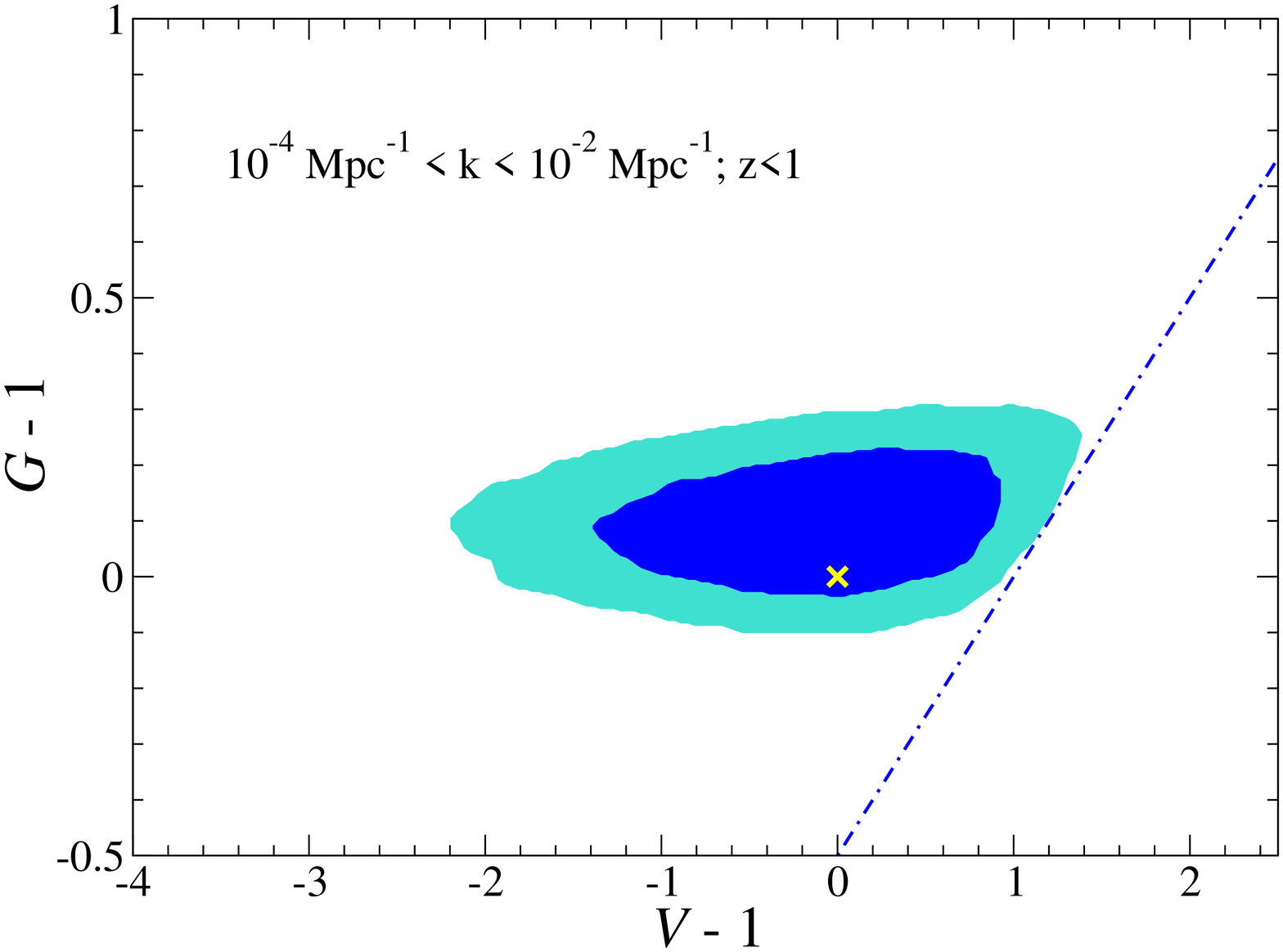}
\label{fig:nogbin2}}
\subfigure[]{\includegraphics[angle=-90,width=\columnwidth]{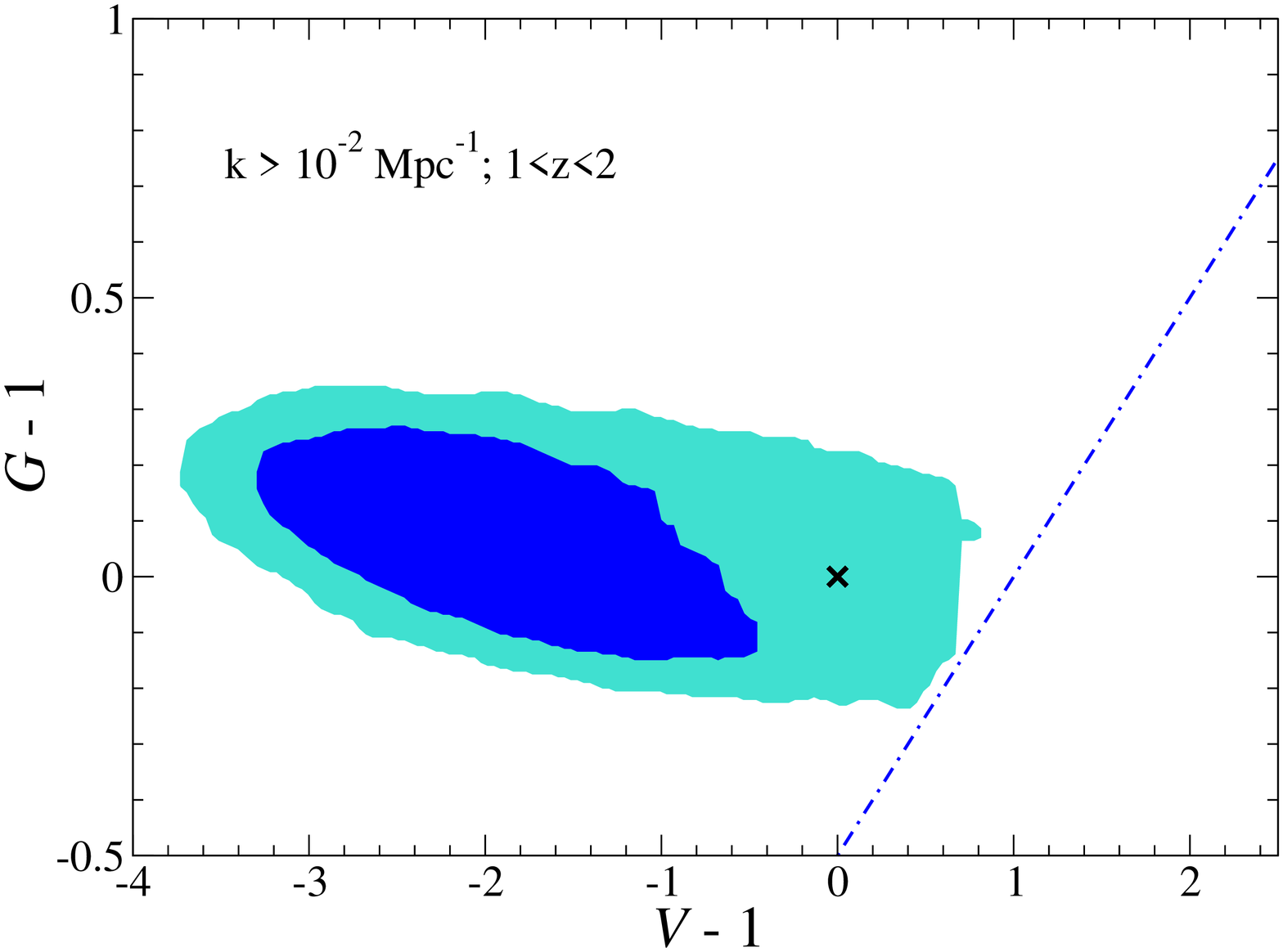}
\label{fig:nogbin3}}
\subfigure[]{\includegraphics[angle=-90,width=\columnwidth=90]{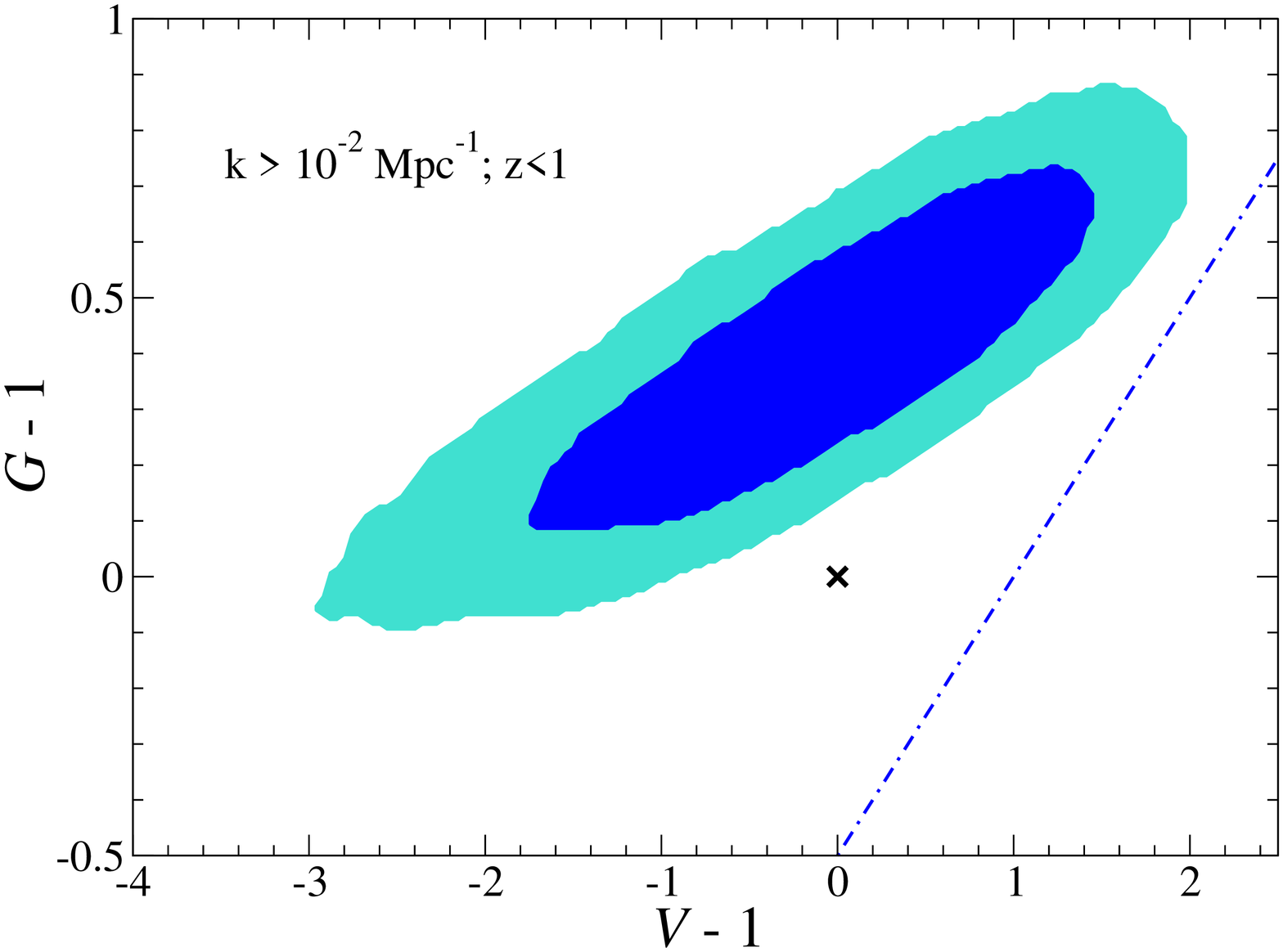}
\label{fig:nogbin4}}
\caption{
68\% and 95\% confidence limit contours for $\vscr-1$ and $\gscr-1$ are 
given for $2\times2$ binning in redshift and $k$ space, using
WMAP7 \cite{Jarosik:2010iu}, Union2 \cite{amanullah}, and
CFHTLS \cite{Fu:2007qq} data.  
The diagonal, dot-dashed line denotes
values of $\vscr$ and $\gscr$ for which $\mu=0$ and gravity vanishes 
(see Eq.~\ref{eq:mudef}).  The x's denote GR values. 
}
\label{fig:nogfig} 
\end{figure*}

In Fig.~\ref{fig:nogfig} we plot the $68\%$ and $95\%$ confidence limit
contours in $\{\vscr,\gscr\}$ space for all bins of $z$ and $k$.  
These contours have been calculated generalizing the modified 
COSMOMC code used in~\cite{gr1} and incorporate
WMAP7 \cite{Jarosik:2010iu}, supernova Union2 \cite{amanullah}, and CFHTLS
weak lensing \cite{Fu:2007qq} data.  The original COSMOMC was presented in 
\cite{Lewis:1999bs,Lewis:2002ah,COSMOMC_notes} and the weak lensing 
data module is from \cite{Lesgourgues:2007te}. 

From Figs.~\ref{fig:nogbin1}-\ref{fig:nogbin3} we see that our initial
supposition that $\vscr$ and $\gscr$ are mostly independent (or, at least, 
less correlated than $\varpi$ and $\mu$ are in Fig.~\ref{fig:muvarpi6}) is 
correct.  We also see that the constraint on $\gscr$ is in all cases stronger
than the constraint on $\vscr$.  

The dot-dashed line on the right side of
the figures corresponds to values of $\vscr$ and $\gscr$ for which $\mu=0$
(see Eq.~\ref{eq:mudef}).  This would imply that the metric is independent 
of matter perturbations, which seems unphysical. 
For the most part, restriction to $\mu>0$ does not strongly affect the 
contours.  However, the data considered so far is not so strong as to 
exclude the $\mu<0$ region (to the right of the line) without imposing 
a prior.  We feel the prior is justified in that, referring back to 
Eq.~(A5) of \cite{gr1}, $\mu<0$ implies (as a consequence of stress-energy 
conservation) that there will be some value of $k$ for which a factor on 
the left hand side of that equation goes to zero, causing $\ddot\Delta_m$ to 
diverge.

The upper two figures show the results for the low $k$ bin, where 
the current data is most constraining.  Note that GR is comfortably 
within the 68\% cl contour.  The constraints above $z\approx1$ are 
slightly tighter, since the ISW effect is more sensitive to this region. 
The bottom two figures give the results for the high $k$ bin, and 
here $\gscr$ is significantly more constrained at higher $z$, again 
due to the ISW.  Comparison of the different $k$ bins in the same 
redshift range show that low $k$ is better constrained, due to the ISW. 

Figure~\ref{fig:nogbin4}, the high $k$ -- low $z$ case, exhibits a number 
of peculiarities.  It is much less constrained than the other cases, and 
shows a higher correlation between $\gscr$ and $\vscr$.   There is also 
an apparent, nearly 3$\sigma$ exclusion of General Relativity. 
Although this is a tantalizing result, it should not be taken too 
seriously.  Since the ISW effect is an integral over redshift, the low 
redshift bins have very weak effects on the CMB anisotropy spectra and 
cannot be tightly constrained by
WMAP7.  Therefore, any systematic errors that create tension between the CFHTLS
data and WMAP will be able to manifest themselves as non-GR values of $\vscr$
and $\gscr$ in these bins.  Furthermore, the high $k$ bins encompass scales for
which the ISW effect is subdominant anyway.  
It behooves us to turn our attention, then, to other data sets that 
may be sensitive to these bins in the hope of strengthening 
our confidence in these constraints.  We do this in the next section. 

Before we proceed, however, we note that this 
is not the first work to find a 2$\sigma$ exclusion of GR
at small scales and low redshift.  
Reference~\cite{Zhao:2010dz} reports a
similar result for their parametrization $\chi_{II}$ (see Section IVB of that
work).  One curious difference, though, is that their analysis shows a 
preference for $\gscr<1$ (their $\Sigma$ is equivalent to our $\gscr$), 
whereas our Fig.~\ref{fig:nogbin4} shows a clear preference for $\gscr>1$.  
This difference can be traced to the different $k$ binning schemes. 
The discussion in~\cite{Zhao:2010dz} attributes the preference for non-GR
$\gscr$ as a means to fit a systematic bump in the CFHTLS weak 
lensing data at large scales 
(see their Fig.~8; they state that the CFHTLS team ascribes 
this to residual systematics; also see Sec.~4.3 of \cite{Fu:2007qq}).  
They divide their $k$ bins at $k=0.1~h~\text{Mpc}^{-1}$.
This is approximately where the bump in the CFHTLS data occurs
(for $z\lesssim1$).  
The MCMC code exploits this by selecting a large value of $\Omega_m$
to increase the overall lensing amplitude and 
fit the bump at large scales (and low $k$) while reducing the value
of $\gscr$ in the small scale (large $k$) bin to prevent
that increased amplitude from spoiling the fit to the smaller scale
data. 
This allows them to 
alter the shape of the weak lensing power spectrum to rise and fall with 
the data.  Since we divide the $k$ bins at $k=0.01~\text{Mpc}^{-1}$, 
however, the same shift in parameters 
would suppress growth, and hence weak lensing power, over 
too large a range of angles.  

\begin{figure}[!t]
\includegraphics[angle=-90,width=\columnwidth]{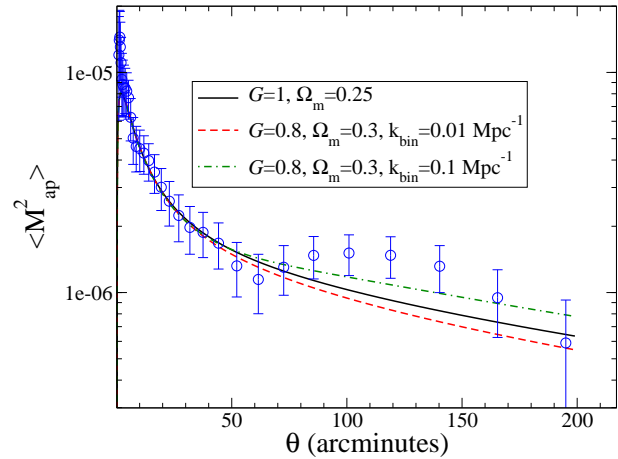}
\caption{
The square of the aperture mass (see Eq.~5 of \cite{Fu:2007qq}) is plotted 
for different cosmological models in comparison to data from
the CFHTLS survey.  The solid, black curve shows the results from the
$\Lambda$CDM concordance model in GR.  One can match the small angle 
behavior by suppressing growth through decreasing the gravitational 
coupling $\gscr$, but increasing growth by increasing $\om$.  
Exploring the larger angular scales, the dashed, red curve 
shows the effect of changing $\gscr(k>0.01\,\text{Mpc}^{-1};z<1)$ while 
compensating $\om$. 
The dot-dashed, green curve shows the case for 
$\gscr(k>0.1~\text{Mpc}^{-1}; z<1)$ as taken in Fig.~8 of \cite{Zhao:2010dz}. 
Because this parametrization divides $k$ bins in the midst of the
scales probed by the data, this curve fits better the (possibly spurious) 
bump in $\langle M^2_{ap}\rangle$ seen in the data between 
$60~\text{arcmin}<\theta<180~\text{arcmin}$.
Data is taken from Table~B2 of \cite{Fu:2007qq}. 
}
\label{fig:zhaomap}
\end{figure}

Figure~\ref{fig:zhaomap} illustrates this, as well as recreating Fig.~8 
of \cite{Zhao:2010dz}.  While the curve that divides $k$ bins at 
$k=0.1~\text{Mpc}^{-1}$ roughly fits the shape of the systematic 
feature between 60 and 180 arcminutes, the curve that uses a division at 
$k=0.01~\text{Mpc}^{-1}$ is actually a worse fit than the GR result.

\begin{figure}[!h]
\includegraphics[angle=-90,width=\columnwidth]{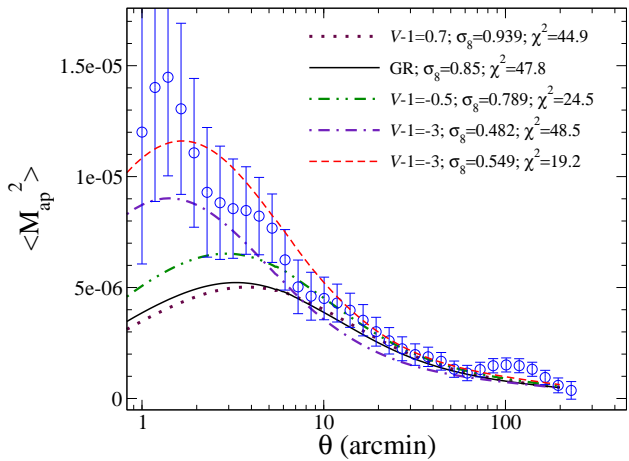}
\caption{A view of the data in Fig.~\ref{fig:zhaomap} with a log 
scale in $\theta$ to zoom in on small angles.  The theory curves use 
bins divided at $k_{\rm bin}=0.01\text{Mpc}^{-1}$, and each is generated 
with identical background cosmology parameters, fixing the amplitude of
the primordial scalar perturbations, so that different post-GR parameter 
values give different values of $\sigma_8$.  Labeled values of $\vscr-1$ 
are set in the high $k$ -- high $z$ bin.  Values of $\gscr-1$ in the
high $k$ -- low $z$ bin are then given by the approximate degeneracy relation 
$\gscr-1=-0.2(\vscr-1)+0.06$.  All other post-GR parameters are set to zero. 
To fit the rise at small angles, much steeper than in GR, requires very 
negative $\vscr$ and hence low $\sigma_8$.  Even raising the primordial 
perturbation amplitude (dashed red curve) cannot bring $\sigma_8$ into 
the usual range. 
Values of $\chi^2$ reported in the legend are calculated naively assuming
a diagonal covariance matrix using the error bars shown.  The four 
smallest-scale data points are excluded from the $\chi^2$ calculation. 
}
\label{fig:ourmap}
\end{figure}

However, the main influence leading to 
our apparent detection of a departure from GR is actually due
to the behavior of the small-angle CFHTLS data (which \cite{Zhao:2010dz}
excludes out of deference to the uncertainties of non-linear modified
gravity), as can be seen in Fig.~\ref{fig:ourmap}.  Each of the curves
in this figure is generated with identical cosmological parameters
($h=0.719$, $\Omega_m=0.25$, and the primordial scalar perturbation 
amplitude, rather than $\sigma_8$, is also fixed; 
an exception to this last rule is
made for the dashed red curve, as discussed in the caption).  
Post-GR parameters are all set to zero
except the high $k$ -- high $z$ value of $\vscr-1$ and the high $k$ 
-- low $z$ value of $\gscr-1$, which are chosen according to the
relationship (which approximately follows the degeneracy direction
of the contours drawn in that parameter space) $\gscr-1=-0.2(\vscr-1)+0.06$.
One sees that decreasing $\vscr-1$ allows the model
to better reproduce the precipitous rise of $\langle M_\text{ap}^2\rangle$
towards small angles.  Such small values of $\vscr-1$ also
reduce the value of $\sigma_8$ predicted.  
The attempt to fit the steep rise in the CFHTLS data not only drives 
$\sigma_8$ down but this in turn then affects other cosmological parameters.

\begin{figure}[!t]
\includegraphics[angle=-90,width=\columnwidth]{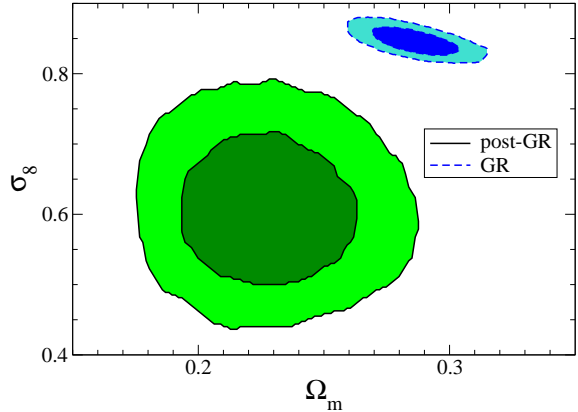}
\caption{The 68$\%$ and 95$\%$ cl contours in $\Omega_m$-$\sigma_8$ space
for WMAP7, Union2, and CFHTLS data in the case of our 
post-GR parametrization (solid, green contour) and in the case of GR 
(dashed, blue contour).
The inclusion of post-GR parameters seems to eliminate the degeneracy
evident in the GR case and pulls the contours to lower values
of $\sigma_8$.  This is due to the influence of the steeply rising small-scale
CFHTLS data, as illustrated in Fig.~\ref{fig:ourmap}.
}
\label{fig:sigomggr}
\end{figure}

Figure \ref{fig:sigomggr}
plots the constraint contours in $\Omega_m$-$\sigma_8$ space both for
post-GR and unmodified GR models.  The freedom in the post-GR 
parameters erases the usual degeneracy between $\Omega_m$ and $\sigma_8$
seen in GR, replacing it with a degeneracy between $\sigma_8$ and our
post-GR parameters, while shifting $\om$.  
Overall, the MCMC code including CFHTLS data is led to prefer much 
smaller values of $\sigma_8$ than
are allowed in GR.  
This extreme shift due to the small angle CFHTLS data, 
and the parametrization-dependence exhibited in 
Fig.~\ref{fig:zhaomap} due to the CFHTLS data bump, give two strong 
reasons to doubt the significance
of the exclusion of GR in Fig.~\ref{fig:nogbin4}.  
Because CFHTLS represents the largest current weak lensing data set, 
we continue to use it in the analysis despite these puzzling behaviors. 
However, we will return to these issues in the next section and see that 
COSMOS weak lensing data (and CFHTLS data above $10'$ with regard to the 
$\sigma_8$ shift) 
does not exhibit these deviations.

\section{Galaxy Auto- and Cross-correlations \label{sec:tggg}} 

In order to get useful constraints on redshift- and scale-dependent 
deviations from GR, we will need to go beyond the basic data sets used 
so far: 
WMAP CMB power spectra \cite{Jarosik:2010iu}, Union2 
supernovae 
distances \cite{amanullah}, and CFHTLS weak lensing \cite{Fu:2007qq}.  In 
particular, different types of cosmological probes, more sensitive to 
density growth, could be useful. 

As discussed in Section II of \cite{Daniel:2009kr}, the CMB anisotropy 
spectrum gives poor constraints on modified gravity.
Note the amorphous, two-lobed shape of the CMB plus supernovae 
contours in Figs.~6-9 
of that work.  This is because the ISW term in the CMB auto-correlation goes
as $(\dot\phi+\dot\psi)^2$ and is thus unaware of sign changes induced
by extreme values of $\gscr$ and $\vscr$ (or in previous works $\varpi$).
The introduction of weak lensing statistics alleviates some of this
uncertainty.  However, much of that ground is lost to the introduction of
the second post-GR parameter (see Fig.~\ref{fig:muvarpi6}). 

To proceed further, we need to include measurements that involve more of 
the interesting physics of modified gravity -- further 
relations between $\phi$, $\psi$, and $\Delta_m$. 
The two probes we add are the cross-correlation of CMB temperature 
fluctuations with the galaxy density field and the auto-correlation of 
the density field, i.e.\ the galaxy power spectrum.

Section II of \cite{Ho:2008bz} discusses the theory of
temperature-galaxy cross-correlations.  See also Section IV of
\cite{Bean:2010zq} for a discussion of how this theory
is altered in non-GR gravity.  The salient point is elucidated
in Eqs.~(4-6) of \cite{Ho:2008bz}: temperature-galaxy
cross-correlations constrain cosmological parameters by
comparing the matter fluctuations $\Delta_m$ traced by the
galaxy distribution with the sources for the metric fluctuations $\phi$ and
$\psi$ responsible for the ISW effect.  Because this ISW effect (not its
auto-correlation in the CMB anisotropy) goes as an integral over redshift
of $\dot\phi+\dot\psi$ times the matter density fluctuation, 
the cross-correlation measurement ends up depending on only one factor of
$\dot\phi+\dot\psi$.  
Thus, these measurements ought to be sensitive to the sign changes
that get hidden in the CMB anisotropy spectrum.

It is even likely that temperature-galaxy (Tg) cross-correlation data 
will meaningfully
constrain $\vscr$, since, at the small scales considered, the 
$$\vscr\equiv\mu(1+\varpi)$$
term in the $\ddot{\Delta}_m$ growth equation (A5) of \cite{gr1} 
becomes dominant, all other modified gravity terms being 
suppressed as $(\mathcal{H}/k)^2$.  

References \cite{Ho:2008bz,Hirata:2008cb} 
provide a module to incorporate cross-correlations of the WMAP temperature maps
with galaxy survey data from
the 2-Micron All Sky Survey, the Sloan Digital Sky Survey, and the
NRAO VLA Sky Survey into COSMOMC.  We modify this module
to accommodate non-GR values of $\vscr$ and $\gscr$ and include it
into our modified COSMOMC.  We excise the module code for incorporating weak 
lensing of the CMB described in \cite{Hirata:2008cb}, so as to obtain a 
clearer picture of the impact of Tg information. 

Several works have already applied Tg correlations to the question of 
constraining modified gravity.  Reference~\cite{Lombriser:2009xg} used Tg 
data to constrain DGP gravity models.  They found, as in \cite{Ho:2008bz}, 
that the principal advantage to this data was in constraining models with 
nonzero $\Omega_k$.  Reference~\cite{Lombriser:2010mp} considered $f(R)$ 
gravity and found significant improvement over previous constraints 
using just the CMB, though they also found that galaxy cluster abundances 
gave constraints that were stronger still. These results would seem to 
indicate that Tg data is not as useful at testing gravity as more direct 
measurements of the matter power spectrum.  

However, these studies were 
carried out in the contexts of specific gravity theories in which the 
relationship between high and low $k$ is forced by the theory.  Since we 
make no such assumption, we expect (and find) that inclusion of the Tg 
data significantly improves constraints on our high $k$ post-GR parameters. 
Indeed, \cite{Bean:2010zq} included Tg data in their exploration of a 
model-independent parametrization based on $\vscr$ (which they call $Q$) 
and $\varpi$ (their $R\equiv 1+\varpi$).  Their parameters exhibited a 
similar degeneracy to that discussed in Sec.~\ref{sec:vmu}, however 
the linearity of Tg correlations 
in $(\dot\phi+\dot\psi)$ still allowed them to place tighter constraints 
on the difference $\vscr-\varpi$ than CMB and supernova data alone (see 
their Table 1 and Figs.~5-6). 

We also include measurements of the galaxy-galaxy (gg) auto-correlation 
power spectrum of luminous red galaxies taken from data release 7 of 
the Sloan Digital Sky Survey and incorporated into COSMOMC by a publicly 
available module \cite{Reid:2009xm}.  These measurements
should principally be sensitive to $\vscr$ since 
they are more directly measurements of $\Delta_m$ than of $\phi+\psi$, 
and they are taken at scales $k\ge0.01~\text{Mpc}^{-1}$.

At small $k$, adding temperature-galaxy (Tg) and galaxy-galaxy (gg) 
correlation data produces little change in the constraints on $\vscr$ 
and $\gscr$ vis-\`a-vis Figs.~\ref{fig:nogbin1} and \ref{fig:nogbin2}.
WMAP constraints from the ISW effect dominate at low $k$, plus there 
are no galaxy-galaxy data points at $k\lesssim0.03~h~\text{Mpc}^{-1}$ 
(see Fig.~8 of \cite{Reid:2009xm}).  

\begin{figure}[!t]
\subfigure[]{
\includegraphics[angle=-90,width=\columnwidth]{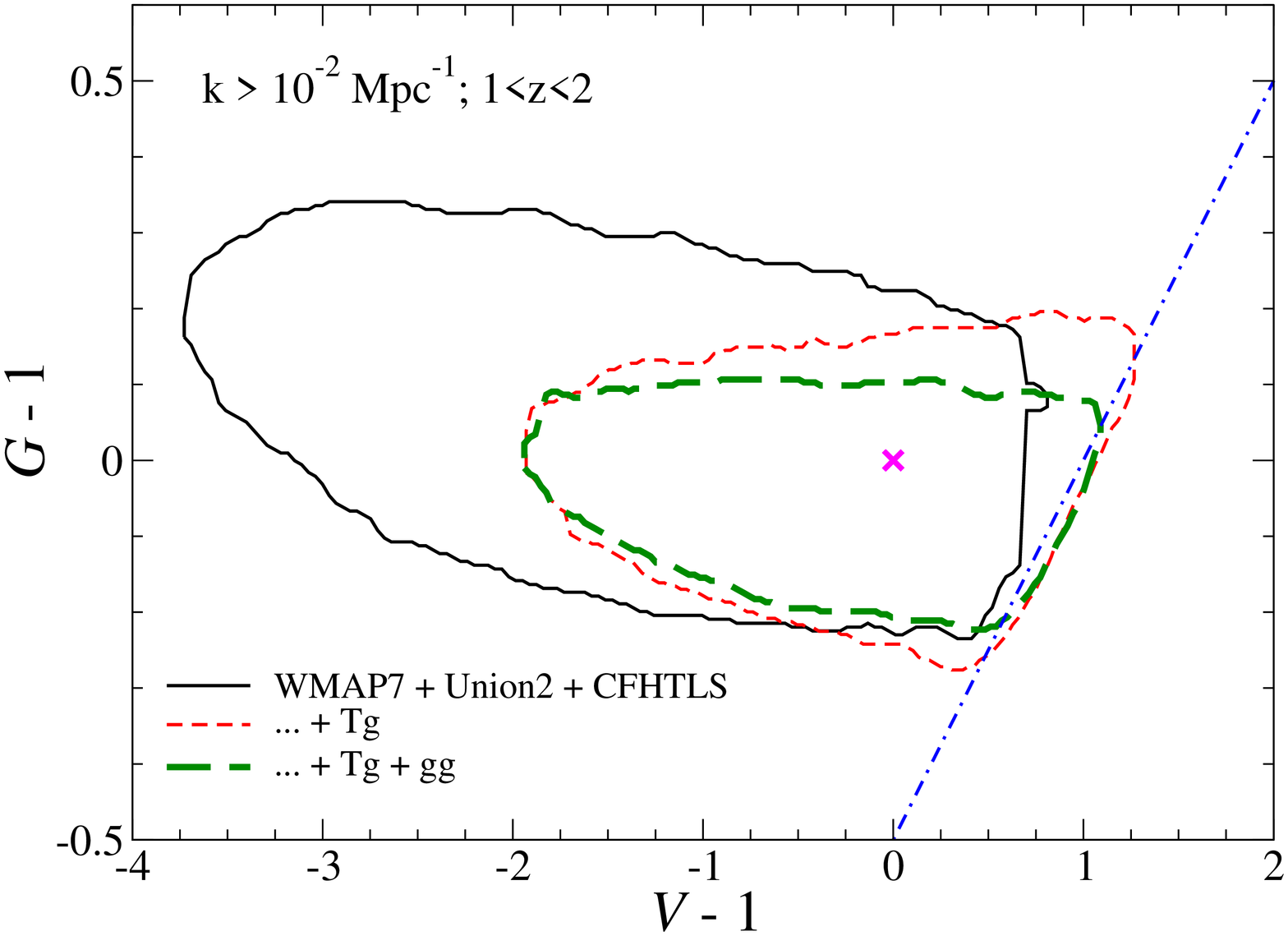} 
\label{fig:bin3comparison}
} 
\subfigure[]{
\includegraphics[angle=-90,width=\columnwidth]{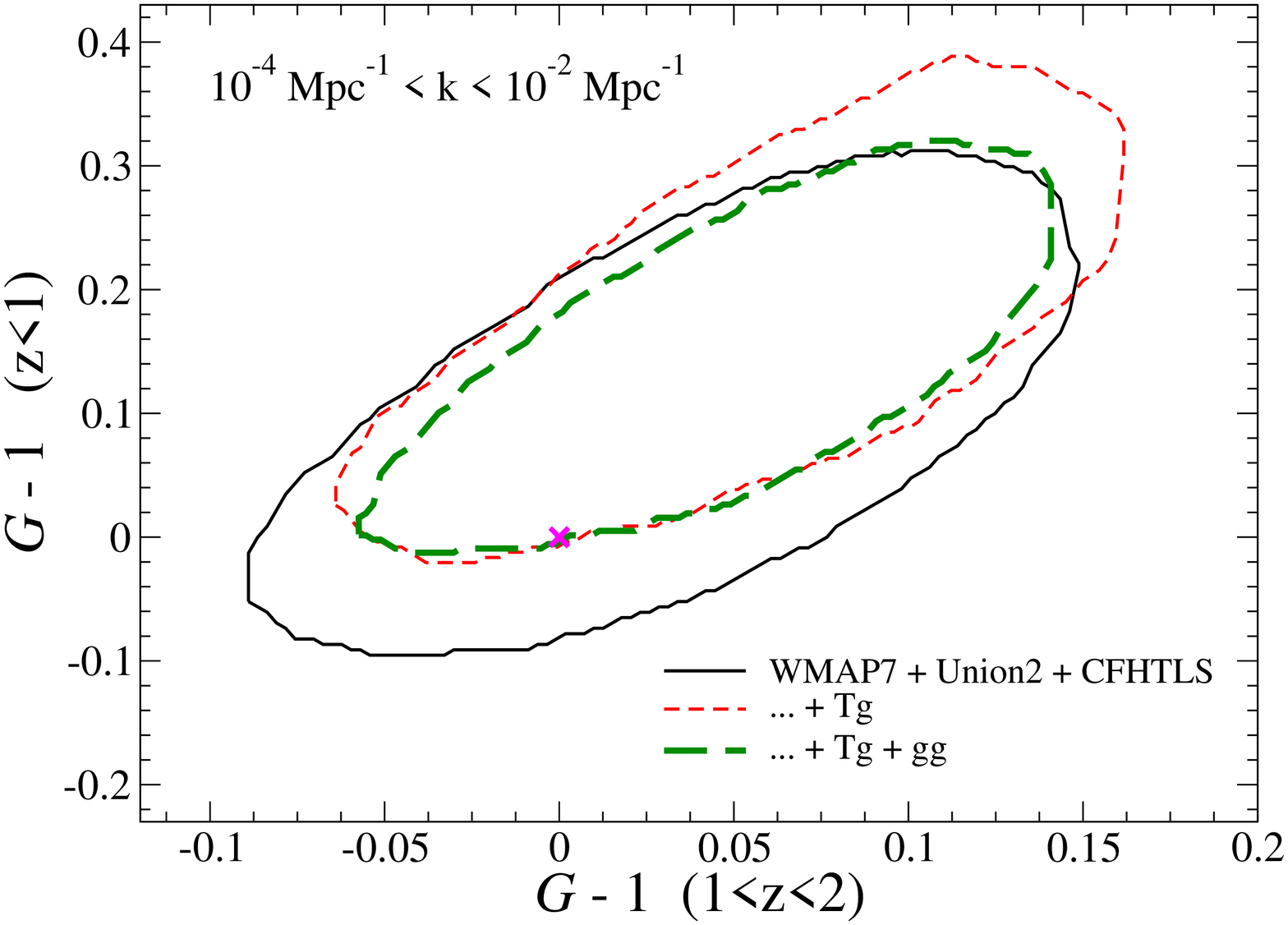}
\label{fig:ggcomparison}
}
\caption{
[Top panel] 
$95\%$ cl contours in $\vscr$-$\gscr$ space for the high $k$, high $z$ 
bin are compared for three different
combinations of data sets.  The solid, black contour shows the results
from Fig.~\ref{fig:nogbin3} using CMB, supernovae, and weak lensing data. 
The thin-dashed, red contour adds temperature-galaxy (Tg) cross-correlation 
data from \cite{Ho:2008bz}.  The thick-dashed, green contour further adds 
galaxy-galaxy (gg) correlation data from \cite{Reid:2009xm}.  The diagonal, 
dot-dashed line gives the $\mu=0$ boundary.  As datasets are added,
the contours close in on GR parameter values (the magenta x).
Current galaxy correlation data is not yet sensitive enough though 
to put a meaningful constraint on $\vscr$. 
[Bottom panel] Addition of data sets can sometimes shift rather than 
tighten the contours, as shown here in $\gscr(1<z<2)$-$\gscr(z<1)$ space 
for the low $k$ bin.  Note that with the additional data GR now lies on 
the edge of the 95\% cl region. 
}
\label{fig:datacomparison}
\end{figure}

For large $k$, however, where the CMB provides generally poor constraints, 
the addition of temperature-galaxy and galaxy-galaxy data can significantly 
alter limits on our post-GR parameters.  
Figure~\ref{fig:datacomparison} illustrates two examples of this.  
Figure~\ref{fig:bin3comparison} shows the strengthening of constraints in 
the 95\% cl contour from Fig.~\ref{fig:nogbin3} upon incorporating as 
well the temperature-galaxy data, and the Tg plus galaxy-galaxy data.  
With current galaxy-galaxy data, most of the improvement is due to Tg, 
but one can anticipate that as larger galaxy surveys including next 
generation surveys are completed then galaxy power spectra will become 
an important ingredient in testing gravity (see Sec.~\ref{sec:fut} for 
future projections).  
In particular, $\vscr$ is still not well determined now. 

Conversely, Fig.~\ref{fig:ggcomparison} shows the effects of adding Tg, 
and then Tg plus gg, can for some variables shift the contours instead of 
tightening them.  This may represent a certain tension between data sets; 
it is interesting to note that Fig.~\ref{fig:ggcomparison} finds 
deviation from GR at the 95\% cl, and we return to the role of CFHTLS 
tension in this below.

\begin{figure*}[!t]
\subfigure[]{\includegraphics[angle=-90,width=\columnwidth]{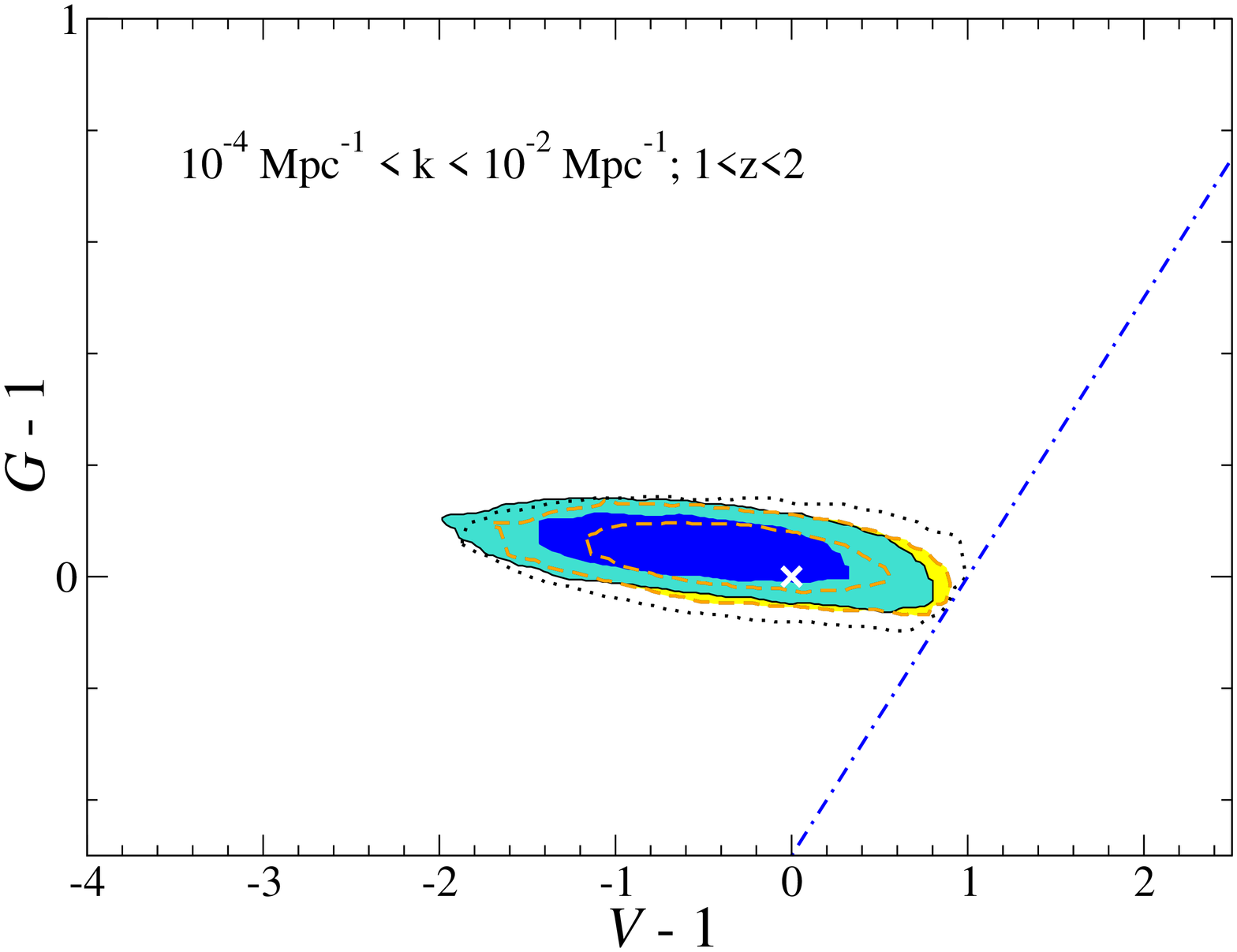}
\label{fig:ggbin1}}
\subfigure[]{\includegraphics[angle=-90,width=\columnwidth]{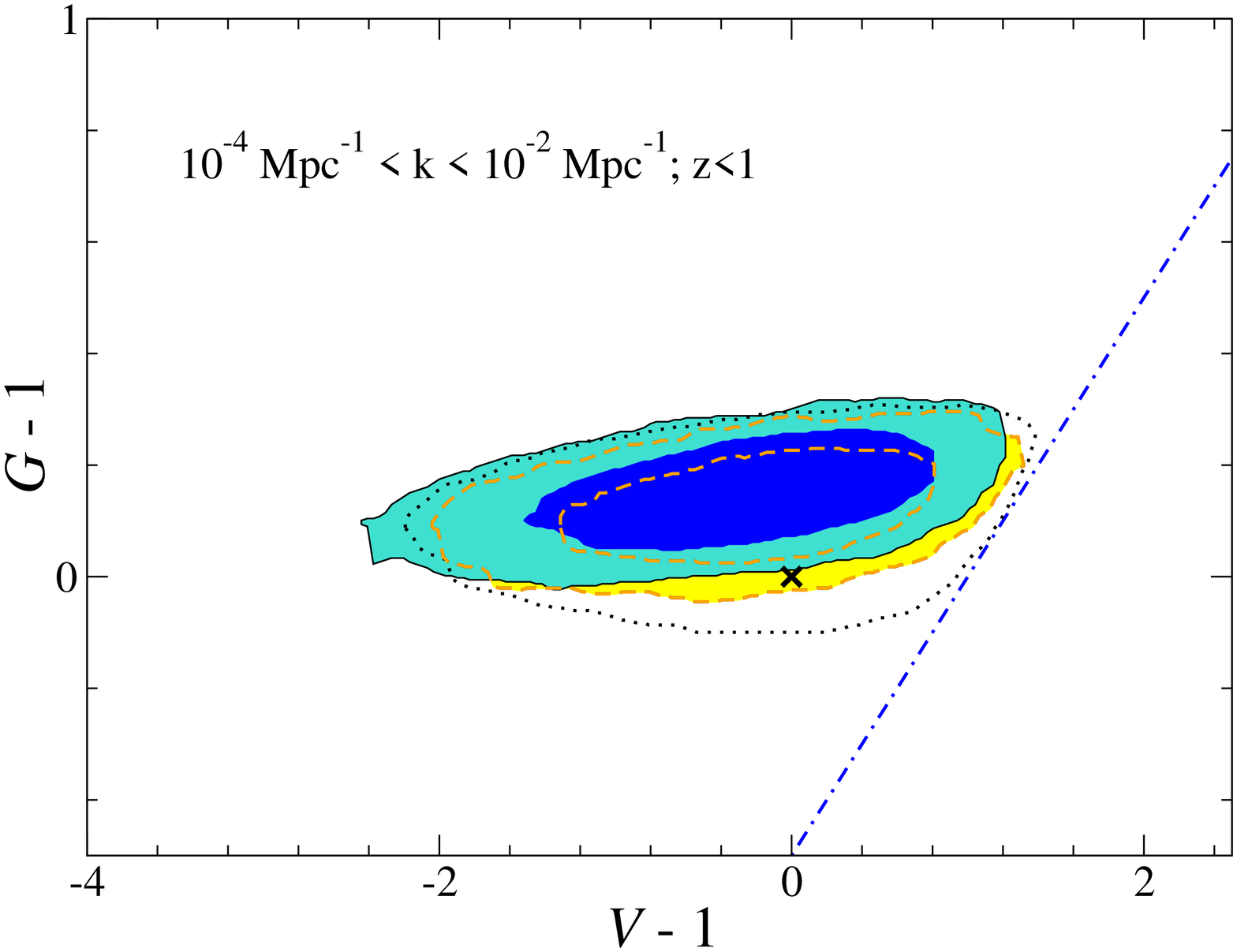}
\label{fig:ggbin2}}
\subfigure[]{\includegraphics[angle=-90,width=\columnwidth]{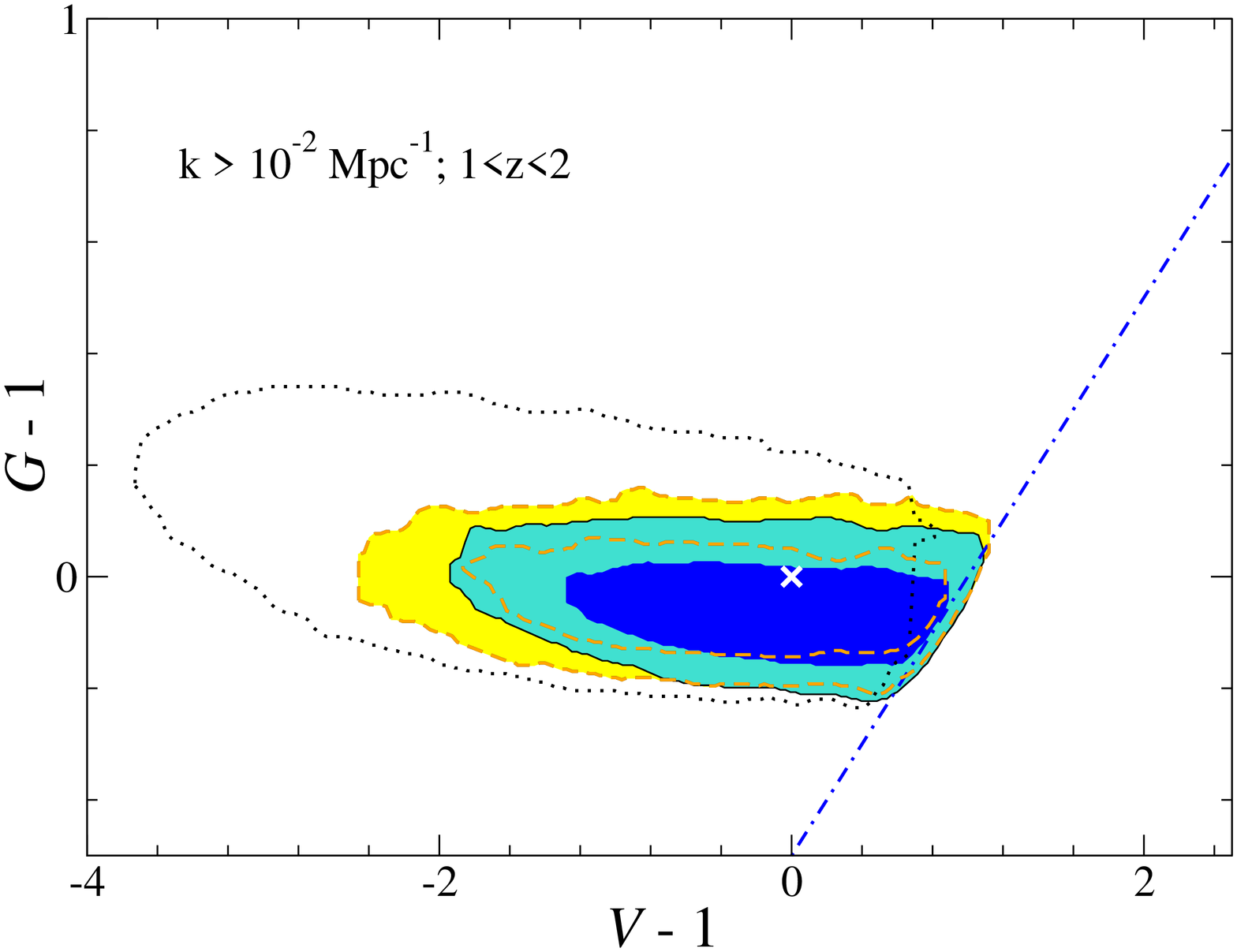}
\label{fig:ggbin3}}
\subfigure[]{\includegraphics[angle=-90,width=\columnwidth]{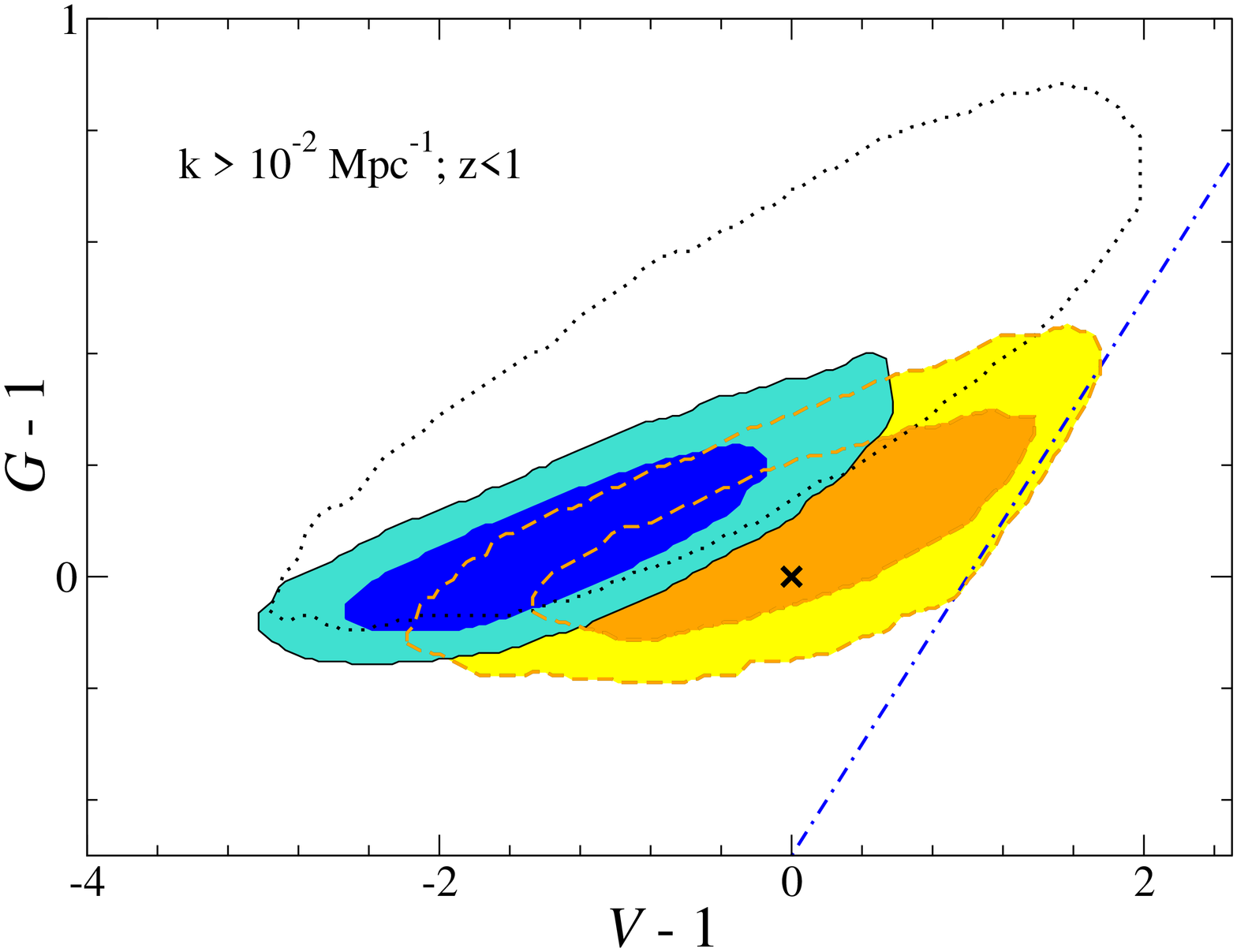}
\label{fig:ggbin4}}
\caption{
68\% and 95\% cl constraints on $\vscr-1$ and $\gscr-1$ are plotted for 
the two redshift and two wavenumber bins, using CMB, supernovae, weak 
lensing, Tg, and gg data.  Foreground, blue contours 
use CFHTLS weak lensing data.
Background, yellow contours use COSMOS weak lensing data.  
The dotted contours reproduce the
95\% cl contours without Tg or gg data from Figs.~\ref{fig:nogfig}.  
The diagonal, dot-dashed line gives the $\mu=0$ 
boundary, from which the low $k$ contours at least have now pulled away.  
The x's denote GR values.  Both $k$ bins at low $z$ exhibit 
some preference for non-GR parameter values when using CFHTLS, but 
not when using COSMOS, weak lensing data.} 
\label{fig:ggfig} 
\end{figure*}

Figures~\ref{fig:ggfig} update all of the plots in Figs.~\ref{fig:nogfig}
using all of the data sets discussed.  The results are the
foreground, blue contours.  We see that $\gscr$ is constrained 
with an uncertainty of roughly 0.1, while $\vscr$ is unknown to within 
$\sim1$.  All cases, except high $k$ -- high $z$, have pulled further off the 
$\mu=0$ restricted area.  Note that the 95\% cl contour in
the high $k$ -- low $z$ bin, Fig.~\ref{fig:ggbin4} using CFHTLS data, still 
excludes General Relativity, although as we have stated this is possibly 
due to systematics in the CFHTLS weak lensing data.  
It now
appears that the low $k$ -- low $z$ bin also prefers non-GR
values of our parameters, though in this case the apparent exclusion
of GR is just at 95\% cl.  
Since this effect did not manifest itself
until we added the galaxy-based datasets, this could either 
be an effect of systematic tension between galaxy-count measurements 
and other data sets, or a true restriction from the increased precision. 
Note that Fig.~\ref{fig:ggcomparison} gives another view 
of the low $k$ -- low $z$ deviation in $\gscr$.  

To test the hypothesis that the exclusion of GR at low $z$
is due to systematic effects in the CFHTLS data, 
we plot the same constraints substituting
weak lensing data from the COSMOS survey \cite{Massey:2007gh} 
in the place of CFHTLS data.  COSMOS constraints
are the background, yellow contours in Fig.~\ref{fig:ggfig}.  
The constraints in the low $k$ bins appear almost unaffected by this
substitution (though the low $k$ -- low $z$ bin no longer hints at an exclusion
of GR, as it did in the case of the CFHTLS data).  This should not be
surprising.  The low $k$ bins correspond to scales where the WMAP7 data has
a lot of constraining power.  The high $k$ -- high $z$
bin also appears moderately insensitive to which weak lensing set is used.  
In the high $k$ -- low $z$ bin, though, we find that 
the 99\% cl exclusion of GR vanishes when COSMOS is used, and GR 
instead lies comfortably within the 68\% cl contour. 
This could mean that the COSMOS
data is less subject to spurious systematic effects.  
Note that the combination using COSMOS data is somewhat less 
constraining due to the small sky area of COSMOS. 
It will be interesting to see what results occur once we have data from 
larger, more detailed future weak lensing surveys.

\begin{figure}[!t]
\subfigure[]{
\includegraphics[angle=-90,width=\columnwidth]{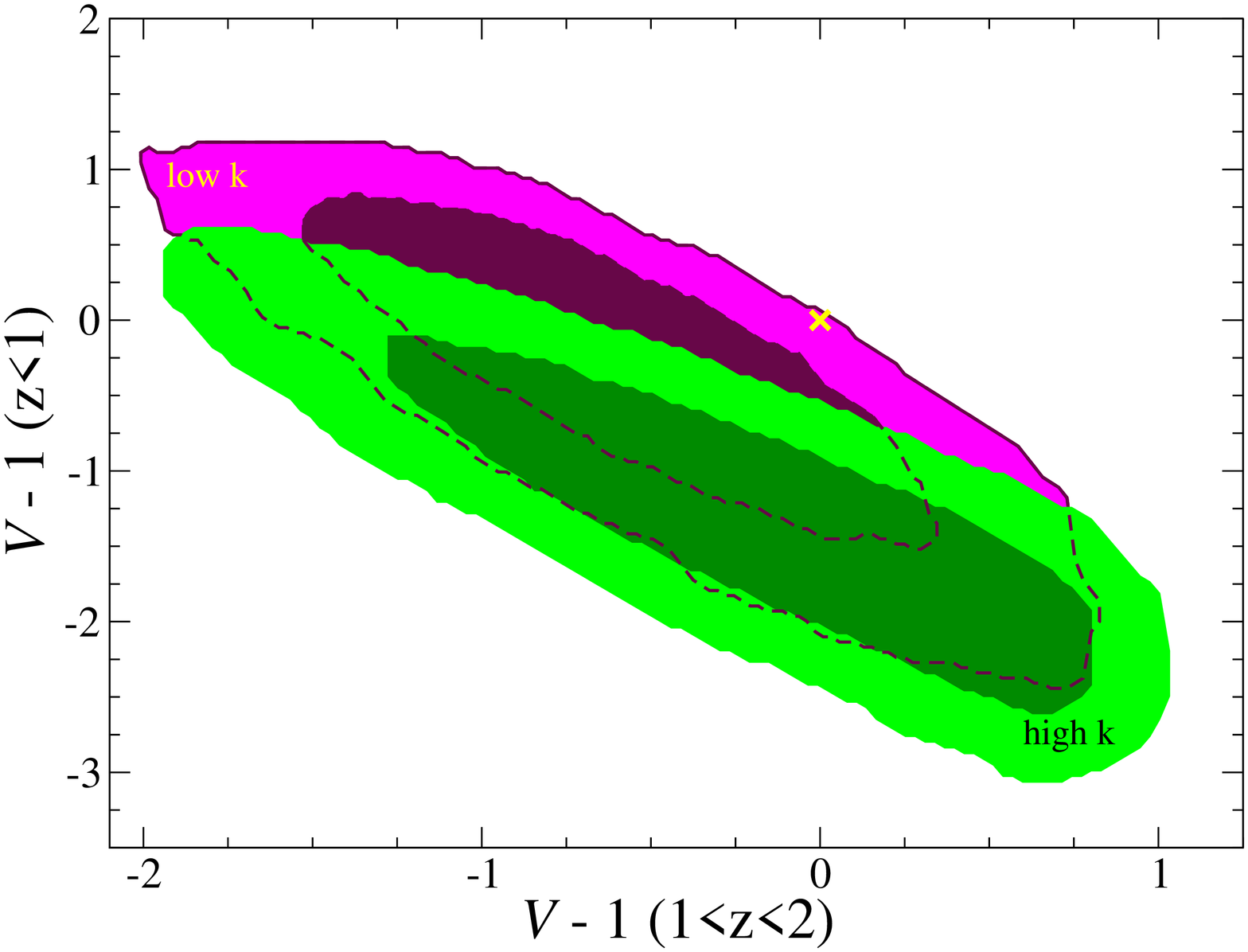}
\label{fig:ggss}
}
\subfigure[]{
\includegraphics[angle=-90,width=\columnwidth]{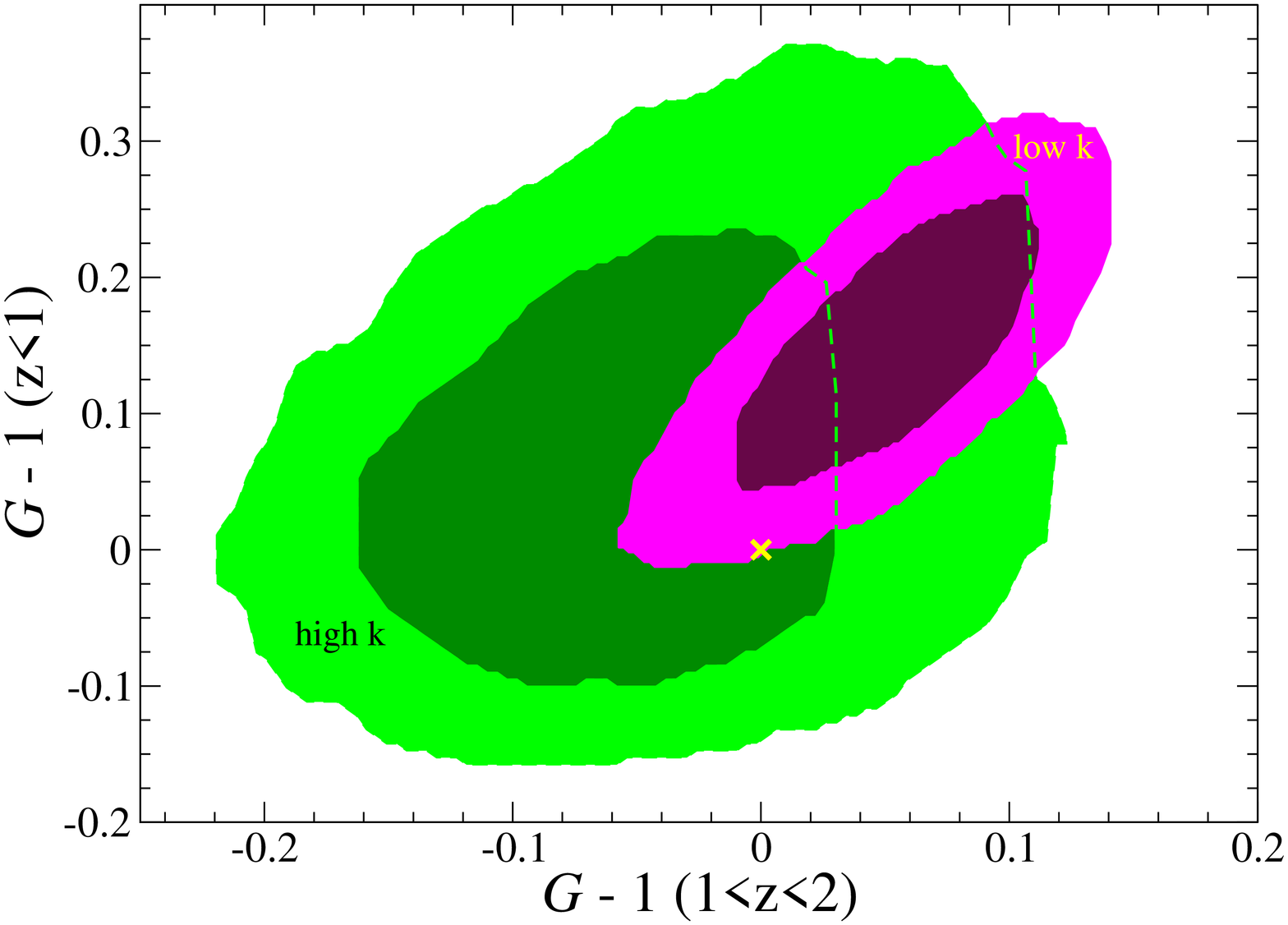}
\label{fig:gggg}
}
\caption{
68\% and 95\% cl constraints on the correlations of $\vscr-1$ (top panel) 
and $\gscr-1$ (bottom panel) between redshift bins, using CMB, supernovae, 
weak lensing (CFHTLS), Tg, and gg data.  Contours are labeled according
to $k$ binning. 
The x's denote GR values.  The high $k$ case of $\vscr$ exhibits a
deviation from GR, corresponding to a different growth amplitude. 
}
\label{fig:ggssfig} 
\end{figure}

\begin{figure}[!t]
\includegraphics[angle=-90,width=\columnwidth]{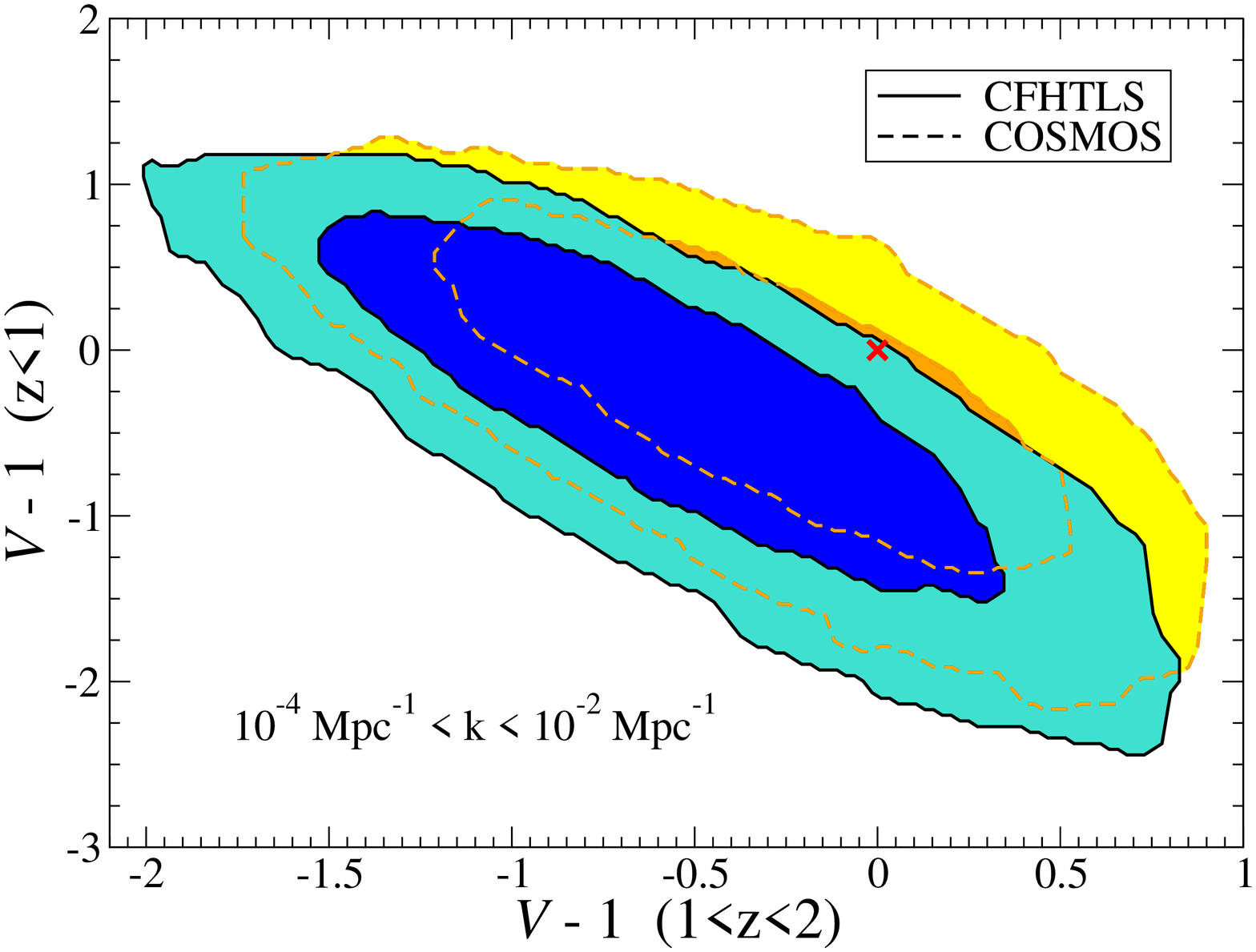}\\ 
\includegraphics[angle=-90,width=\columnwidth]{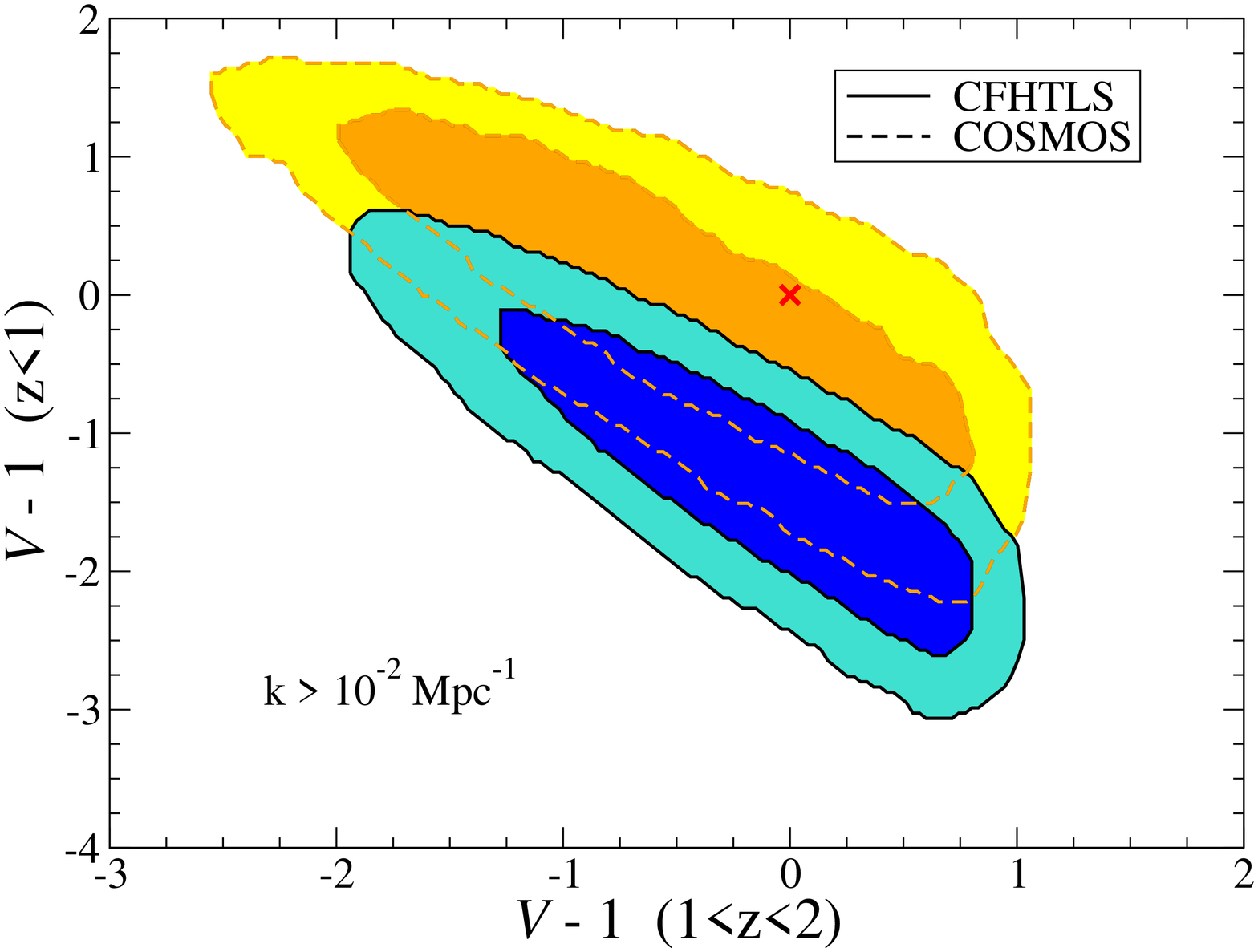}\\ 
\includegraphics[angle=-90,width=\columnwidth]{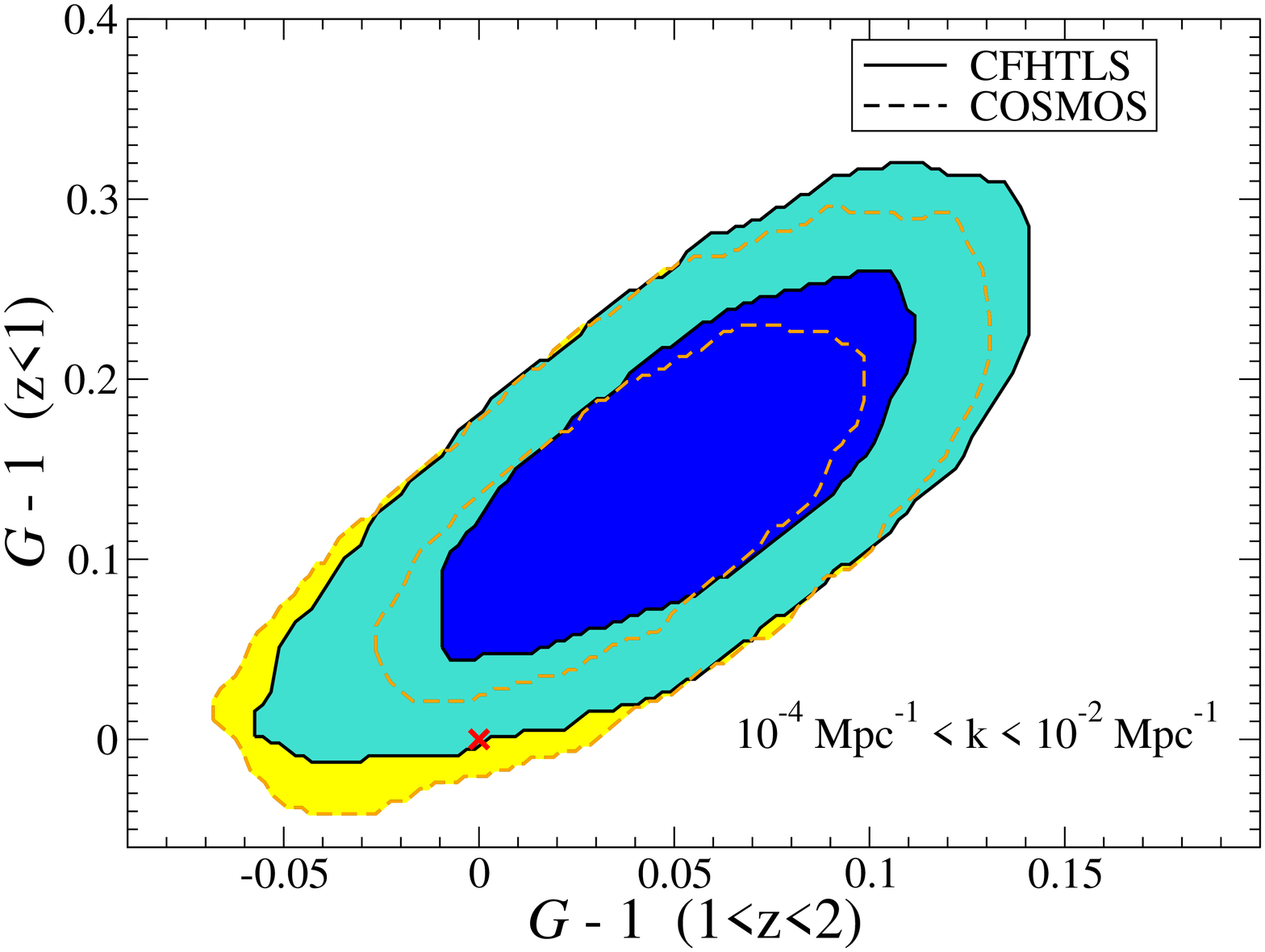}
\caption{
68\% and 95\% cl constraints on the correlations of
$\vscr-1$ and $\gscr-1$ between redshift bins, using CMB, supernovae, weak 
lensing, Tg, and gg data.  Substituting COSMOS (background) 
for CFHTLS (foreground) eliminates 
the apparent exclusion of GR.  The x's denote GR values.  
We do not plot the high $k$ bin correlations for $\gscr-1$ as there are 
no interesting correlations or deviations there.  
}
\label{fig:sscosfig} 
\end{figure}

We can also reexamine the redshift dependence of each of the post-GR 
parameters.  Figures~\ref{fig:ggssfig} plot the correlations between 
redshift bins for $\vscr$ and $\gscr$, for each bin of wavenumber $k$
using CFHTLS data.
The values of $\gscr$ for low $k$ are positively correlated, as found 
before for scale-independent $\gscr$, while for high $k$ the values are 
rather independent.  This can be 
understood by considering that the small $k$ bin has the greatest effect on the
ISW imprint in the CMB.  Since the ISW effect goes as $(\dot\phi+\dot\psi)^2$,
the CMB data will prefer parameter combinations that minimize the change in
$\gscr$ across redshift bins.  Because the large $k$ bin encompasses scales
over which the ISW effect is subdominant, such a preference is not operative 
there.

For $\vscr$ there is negative correlation as before, for both low and high 
$k$.  This is a manifestation of the role of $\vscr$ in regulating the 
growth of structure, i.e.\ $\Delta_m$.  The low $k$
contours in Fig.~\ref{fig:ggss}
have a main degeneracy direction parallel to lines 
with a slope in $\{\vscr(1<z<2),\vscr(z<1)\}$ space of $\sim -1.3$.
The high $k$ contours have a degeneracy direction of 
approximate slope $\sim-1.2$. 
Integrating Eq.~(A5) of \cite{gr1} for values of $\Omega_m$ close to
those favored by WMAP and Union2 ($\Omega_m\sim0.25$), one finds that 
(independent of $\gscr$ and $H_0$) values of $\{\vscr(1<z<2),\vscr(z<1)\}$
that lie along lines with slopes ranging between
$-1.3$ and $-1.4$ return values of the relative growth  
$\Delta_m(z=0)/\Delta_m(z=100)$ that are similar to within a few percent.
Displacement perpendicular to this direction controls the absolute 
growth factor.  The offset of the high $k$ contours
from the low $k$ contours (and from GR), signifies a preference for
suppressed growth relative to GR in the high $k$ modes; GR lies outside 
the 95\% cl contour in the high $k$ bin.  
Again, this could be due to the odd bump in the CFHTLS weak 
lensing power or the steep rise towards small angles in 
Fig.~\ref{fig:ourmap}.  
Figures~\ref{fig:sscosfig} show the effect of using COSMOS weak lensing
data instead of CFHTLS data on the correlations of $\vscr-1$ and $\gscr-1$ 
across redshift bins.  As before, the exclusion of GR vanishes.

As in the scale-independent case, we find that our post-GR parameters
correlate most strongly with $\sigma_8$ out of all of the usual
cosmological parameters.  Once again, larger $\vscr$ amplifies growth 
and induces a
larger $\sigma_8$ while larger $\gscr$ brings lower values of $\sigma_8$
into agreement with the data.  
These correlations are only manifest in the high $k$ bin, indicating that
they are principally dependent on the weak lensing, Tg, and gg data sets.
The correlation with $\vscr$ appears in the high $z$ bin, as expected 
due to the cumulative effect of growth over time (see discussion in 
Sec.~\ref{sec:zkdep}).  
In the case of $\gscr$, probing the potentials, the correlation with 
$\sigma_8$ appears in the low $z$ bin, as expected for the weak lensing data 
weighted toward $z<1$.  
The shift to low $\sigma_8$ as seen in Fig.~\ref{fig:sigomggr} does not 
occur when COSMOS data is used, or when CFHTLS data is restricted to 
$\theta>10'$.

\section{Constraints Possible with Future Data \label{sec:fut}}

Given the weak constraints on $\vscr$ in Fig.~\ref{fig:ggfig}, 
and possible hints of deviations from GR, it is important to investigate 
the capabilities of future galaxy surveys.  These should provide 
us with more direct measurements of, and much better precision on, the 
growth of density perturbations through the galaxy power spectrum, 
testing GR and improving our knowledge of post-GR parameters.  

We consider the specific example of BigBOSS \cite{Schlegel:2009uw}, 
a proposed ground-based survey intended to constrain cosmology by measuring 
the baryon acoustic oscillations and redshift space distortions in the 
galaxy distribution.  Reference~\cite{Stril:2009ey} explored BigBOSS 
tests of gravity (and dark sector physics) in terms of the gravitational 
growth index $\gamma$ \cite{groexp}, using a Fisher matrix calculation. 
Here, we carry out a more sophisticated Markov Chain Monte Carlo fit to 
simulated data, allowing for scale- and time-dependence in the gravitational 
modifications through our binned $k$, binned $z$ post-GR parameters. 

We generate our mock BigBOSS data around the $\Lambda$CDM, GR 
($\vscr=\gscr=1$) maximum likelihood cosmology of WMAP7. 
The data is considered in the form of measurements of the
redshift-distorted galaxy-galaxy power spectrum
\begin{equation}
P_g(\vec{k};z)=P_g(k,\mu;z)=(b+f\mu^2)^2 \,P(k;z)\,, 
\end{equation}
where $P$ is the power spectrum of matter overdensities, $b$ is the 
linear bias relating the overdensity of galaxy counts to the overdensity 
of matter, $f=d\ln\Delta_m(k,a)/d\ln a$ is the growth factor, and $\mu$ is the 
cosine of the angle $\vec{k}$ makes with the line of sight.  

We take survey parameters, including galaxy number densities, from 
\cite{schlegelmarseille}, and consider emission line galaxies (ELG) and 
luminous red galaxies (LRG) as two separate data sets.  The bias $b$ 
is a function of redshift, 
\begin{equation}
b(z)=b_0\,\frac{\Delta_m(k,z=0)}{\Delta_m(k,z)} \,, \label{eq:bias} 
\end{equation}
where $b_0$ is a nuisance parameter to be marginalized over for each 
data set.  
The fiducial values are $b_{0,\text{ELG}}=0.8$ and $b_{0,\text{LRG}}=1.7$.   
The values and redshift dependence are good fits to current galaxy 
observations, and can be motivated by comoving clustering 
models \cite{nikhil}.  

In calculating the galaxy power spectrum we consider
modes $10^{-4}<k<10^{-1} \text{Mpc}^{-1}$.  As in
\cite{Tegmark:1997rp}, we assume that the covariance matrix of $P_g$ is
diagonal with
\begin{eqnarray}
C_{ij}&\approx& 16\pi^3\delta_{ij}\delta_{mn}
\frac{P_g(k_i,\mu_m)P_g(k_j,\mu_n)}{V_0 d^3k}\nonumber\\
&&\times\left[\frac{1+\bar{n}P_g(k_i, \mu_m)}{\bar{n}P_g(k_i,\mu_m)}\right]^2
\end{eqnarray}
where $V_0$ is the real space volume of the survey and $\bar{n}$ is the
selection function of the survey as a function of comoving distance $r$.
Thus, the likelihood $\mathcal{L}$ 
for a cosmological model which predicts galaxy-galaxy
power spectrum $\hat{P}_g(k,\mu)$ is given by
\begin{eqnarray}
-2\ln[\mathcal{L}]&=&\frac{1}{4\pi}\int dz\frac{dr}{dz}r^2\int d\ln k\,k^3
\nonumber\\
&&\times\int d\mu \frac{(\hat{P}_g-P_g)^2}{P_g^2}
\left(\frac{\bar{n}(r)P_g}{1+\bar{n}(r)P_g}\right)^2
\end{eqnarray}

We also include mock Planck CMB data generated with the COSMOMC module 
FuturCMB \cite{Perotto:2006rj,futurcmb} and mock future supernova data 
based on a space survey of 1800 supernovae out to $z=1.5$ (``JDEM'') 
in our projected data MCMC calculation.  For computational efficiency 
we do not consider the gravitational lensing of the CMB.

\begin{figure*}[!t]
\subfigure[]{\includegraphics[angle=-90,width=\columnwidth]{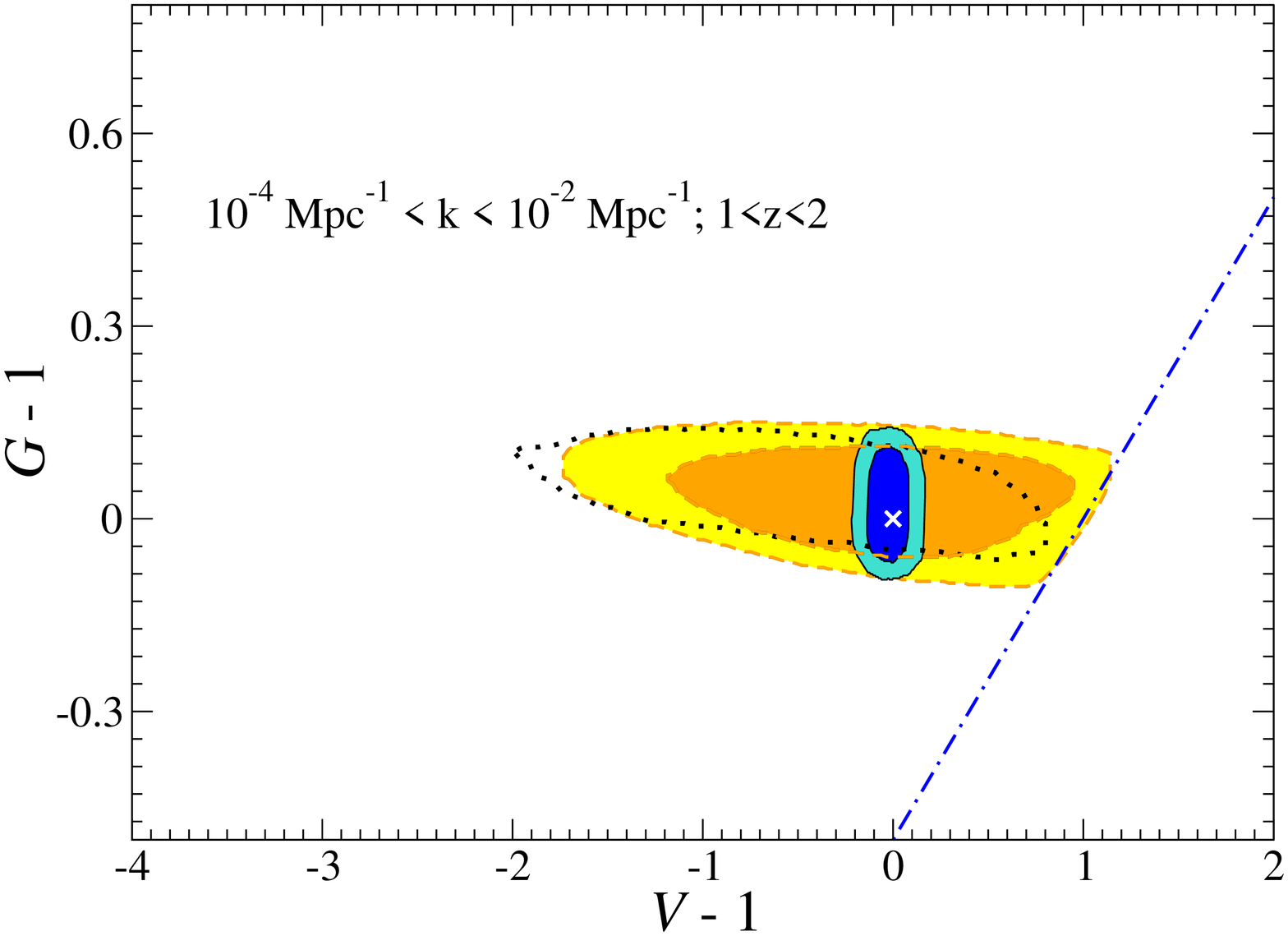}
\label{fig:bbbin1}}
\subfigure[]{\includegraphics[angle=-90,width=\columnwidth]{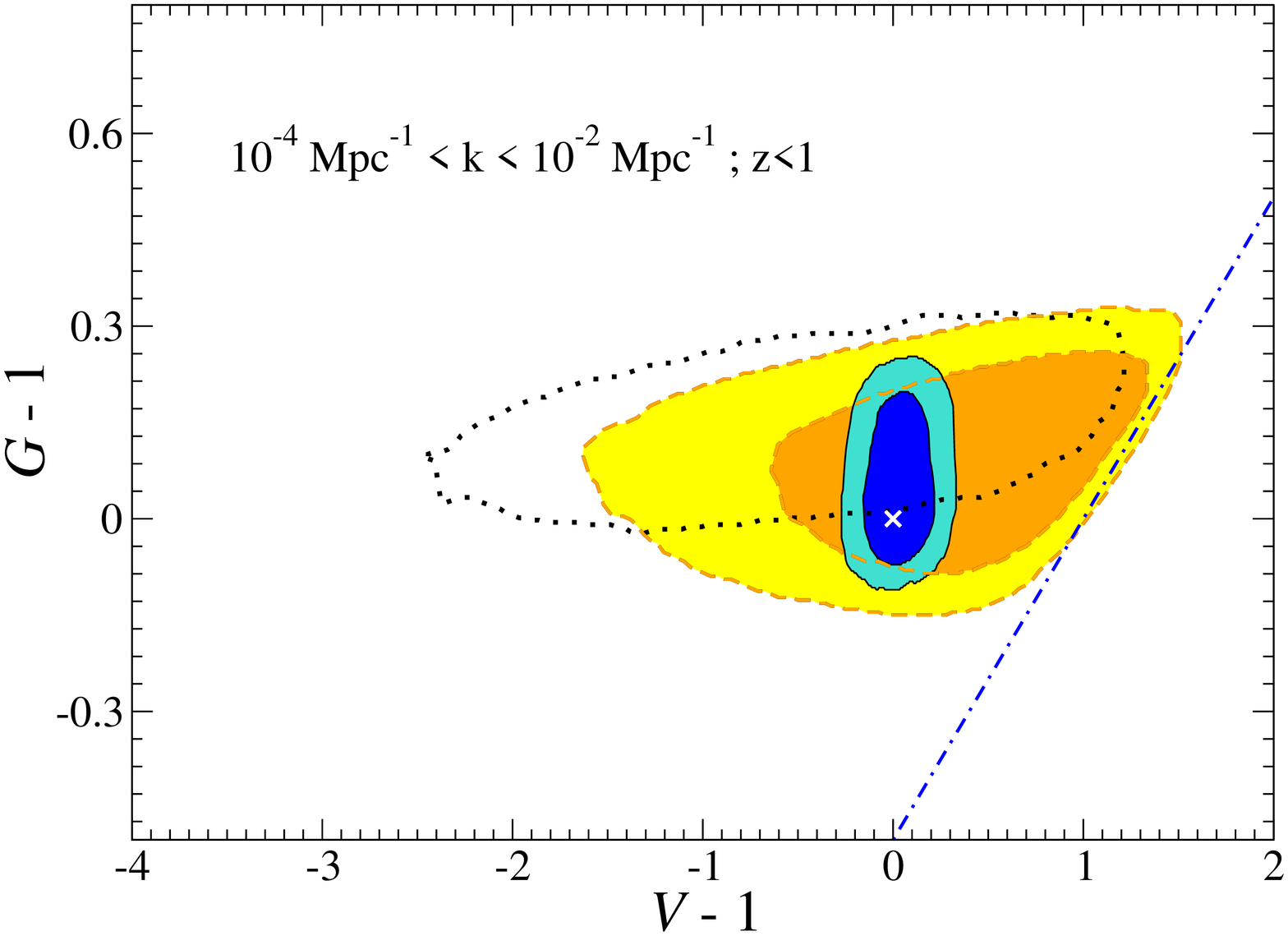}
\label{fig:bbbin2}}
\subfigure[]{\includegraphics[angle=-90,width=\columnwidth]{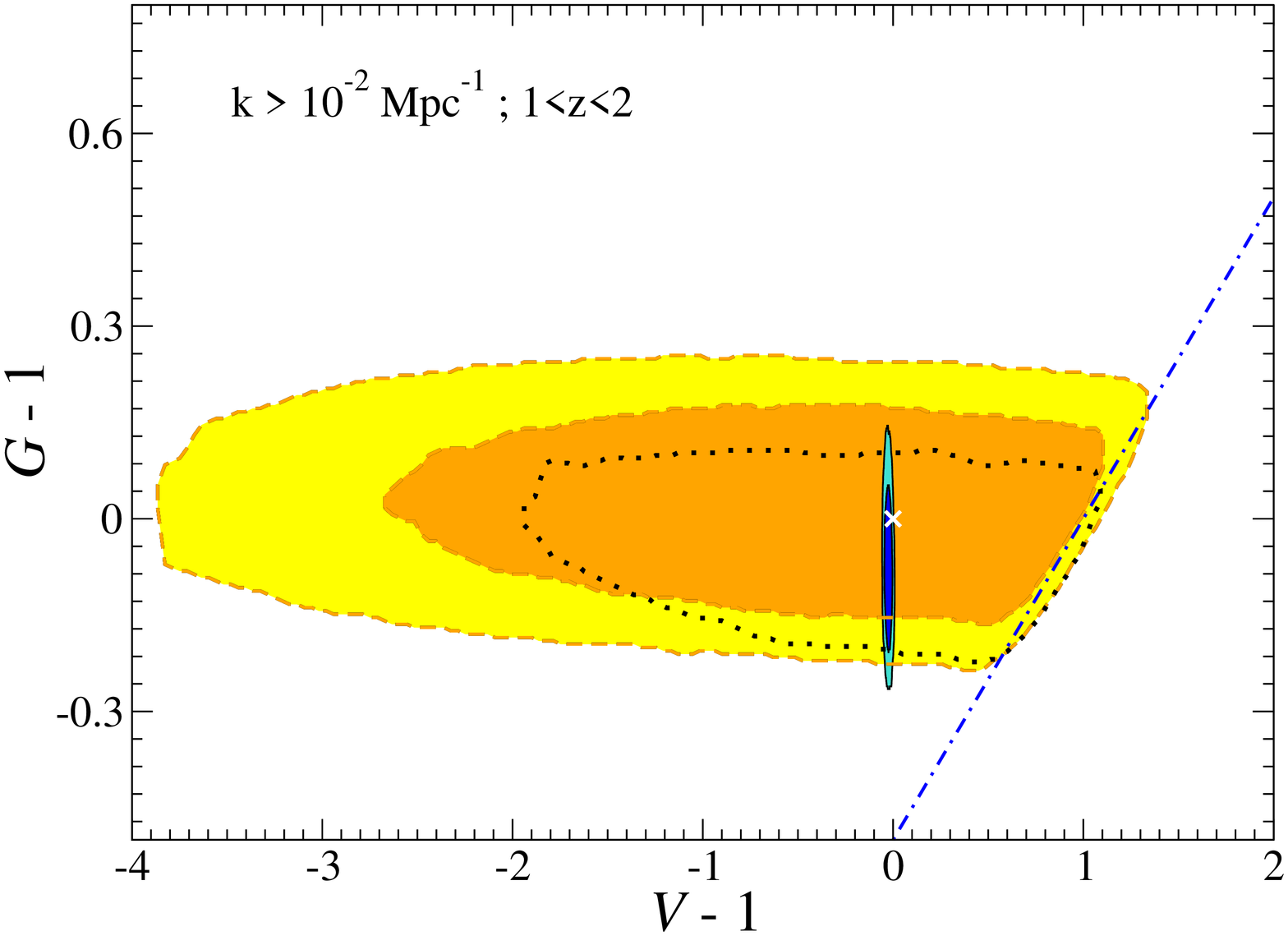}
\label{fig:bbbin3}}
\subfigure[]{\includegraphics[angle=-90,width=\columnwidth]{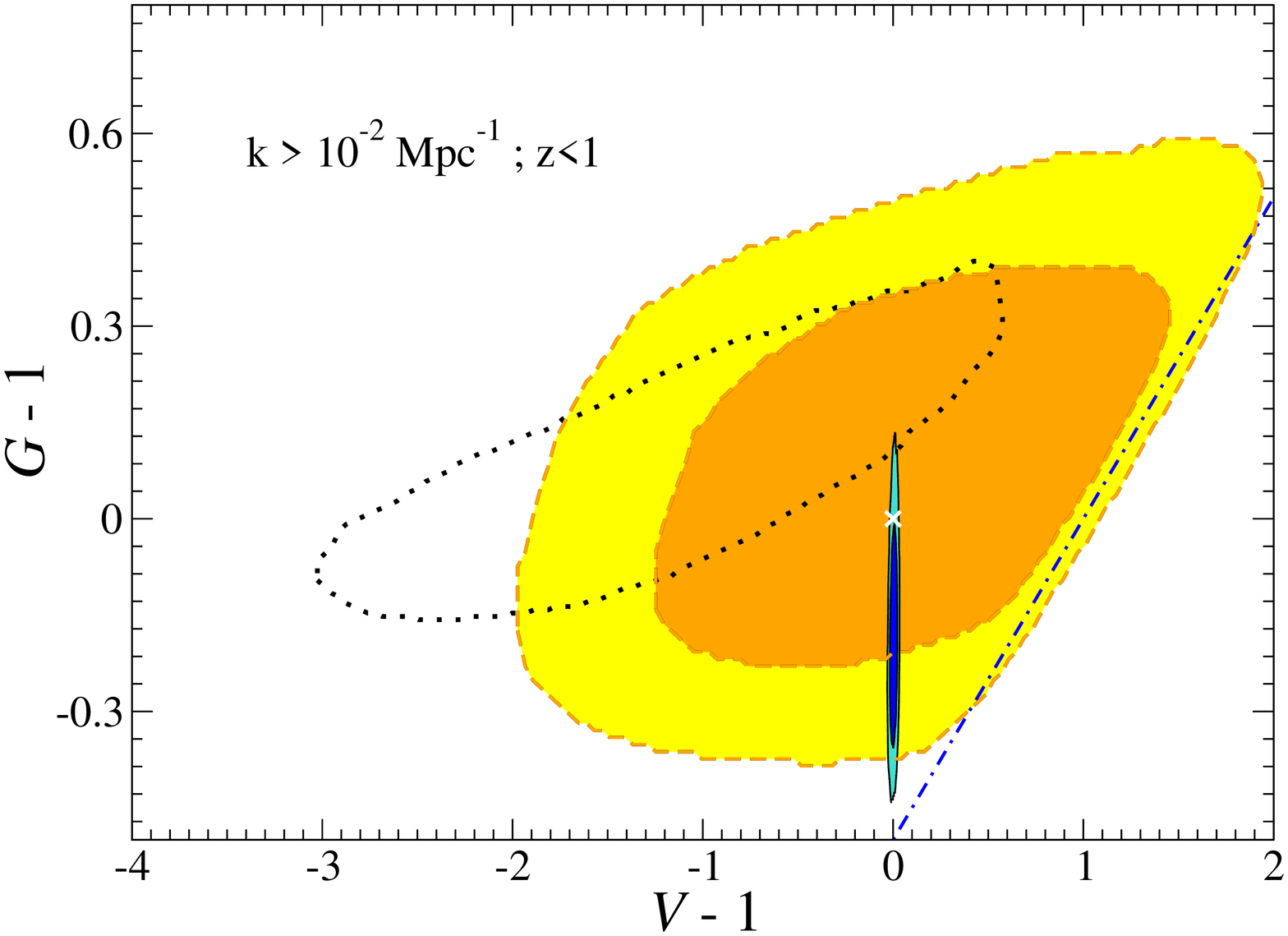}
\label{fig:bbbin4}}
\caption{
68\% and 95\% cl constraints on $\vscr-1$ and $\gscr-1$ are plotted for 
the two redshift and two wavenumber bins using mock future data.
Foreground (blue) contours use mock BigBOSS, Planck, 
and JDEM supernova data.  Background (yellow) contours
use only mock Planck and JDEM supernova data. The dotted contours recreate the
95\% cl current data contours from Figs.~\ref{fig:ggfig} (using CFHTLS) 
to illustrate the expected improvement in constraints.  
The x's denote the fiducial GR values.} 
\label{fig:bbfig} 
\end{figure*}

Figures \ref{fig:bbfig} show the $68\%$ and $95\%$ confidence limit contours
resulting from the full MCMC calculation on our mock data sets.  There is an
order of magnitude or more improvement in the constraint placed on $\vscr$ 
in all bins by including 
BigBOSS.  As in \cite{Stril:2009ey}, we find that BigBOSS will be able to
constrain departures from the growth history of GR (here parametrized as
$\vscr-1$, there by $\gamma$) to within $\sim10\%$.  Constraints on $\gscr$ 
also improve, though by more modest factors.

\section{Conclusions \label{sec:concl}} 

The suite of current cosmological data has grown to the point that 
increasingly sophisticated model-independent extensions to 
general relativity can be tested.  This includes both time- and 
scale-dependent modifications; we utilize bins in redshift $z$ and 
wavemode $k$ for localization of the effects and clarity of physical 
interpretation.  The functions $\gscr$ and $\vscr$ investigated 
here in detail, giving a complete model-independent description 
(together with stress-energy 
conservation) of the gravitational modifications, are closely tied to 
the sum of the metric potentials and to the matter growth, respectively. 
They also have the virtue of being substantially decorrelated 
from each other.  On the other hand, correlations across redshift or 
across length scales can be easily studied, giving deeper insight 
into the effects of the modifications and where they show up in the 
observations. 

Using current CMB, supernova, and weak lensing data from CFHTLS we 
find an inconsistency with general relativity at near the 99\% 
confidence limit at $k>0.01\,{\rm Mpc}^{-1}$ and $z<1$.  Through a 
series of investigations we identify its origin as being due to an 
abnormally steep rise in the weak lensing power at small scales. 
This rise strongly shifts $\vscr$ from the GR value and also drives 
down the estimated value of $\sigma_8$.  
The CFHTLS data also shows an unusual bump in the power at larger scales. 
By replacing the CFHTLS measurements 
with COSMOS weak lensing data, we find that all these deviations 
vanish and that GR provides an excellent fit.  Thus, the deviations 
may originate in systematic effects in the CFHTLS data (a possibility 
also raised by members of the CFHTLS team) interacting with increased 
freedom from the post-GR parameter fitting. 

The addition of galaxy clustering measurements, through both the 
CMB temperature-galaxy count (Tg) and galaxy-galaxy power spectrum (gg) 
statistics, tightens the constraints on smaller scales (high $k$). 
This improvement is especially noticeable in $\vscr$, since it enters 
in the matter growth.  Again, the full combination using CFHTLS data 
shows significant deviations, which go away on substitution with COSMOS 
data. 

Given the important role of galaxy survey data, and the still weak 
constraints on the $\vscr$ deviation parameter (only of order unity), 
we examine the potential leverage of future galaxy survey measurements, 
specifically from BigBOSS.  These appear quite promising for confronting 
general relativity with further measurements, giving a direct probe 
of growth and one that could be highly precise from the large statistics. 
Together with Planck CMB and supernovae data, such a galaxy survey could 
improve the 
area uncertainty on the post-GR parameters in each of four bins of 
redshift-scale by factors from 10 to 100.  This is an exciting prospect 
as we seek to understand gravity as the most pervasive and dominant 
force in the universe.

\acknowledgments

We thank Tristan Smith for helpful discussions and insight and 
Chanju Kim for timely hardware fixes. 
We acknowledge use of NASA's Legacy Archive for Microwave Background
Data Analysis (LAMBDA).  
This work has been supported by the World Class University grant 
R32-2009-000-10130-0 through the National Research Foundation, Ministry 
of Education, Science and Technology of Korea. 
EL has been supported in part by the
Director, Office of Science, Office of High Energy Physics, of the
U.S.\ Department of Energy under Contract No.\ DE-AC02-05CH11231.


\end{document}